\documentclass[useAMS,usenatbib]{mn2e}
\usepackage{amssymb}
\usepackage[pdftex]{graphicx}
\usepackage[pdftex,usenames,dvipsnames]{xcolor}
\usepackage{natbib}
\usepackage{aas_macros}
\usepackage{url}
\usepackage[none]{hyphenat}
\usepackage{times}
\usepackage[linktocpage=true,backref=true,pagebackref=false,bookmarks=true,bookmarksnumbered=False,colorlinks=true,linkcolor=NavyBlue,citecolor=NavyBlue,urlcolor=NavyBlue]{hyperref}
\usepackage[all]{hypcap}
\usepackage[backgroundcolor=yellow,linecolor=red,bordercolor=yellow,textsize=footnotesize]{todonotes}

\pdfpageattr {/Group << /S /Transparency /I true /CS /DeviceRGB>>} 

\def\oiii{\hbox{[O\,{\sc iii}]}}
\def\nii{\hbox{[N\,{\sc ii}]}}

\def\kms{\hbox{km s$^{-1}$}}

\title[Warm molecular gas]{The warm molecular gas and dust of Seyfert galaxies: \\two different phases of accretion?}
\author[Mezcua et al.]{M. Mezcua$^{1,2,3}$, M. A. Prieto$^{2,3}$, J.A. Fern\'andez-Ontiveros$^{4,5}$, K. Tristram$^{6}$, N. Neumayer$^{7}$ \and and J. K. Kotilainen$^{8}$ \vspace{0.2cm}\\
$^1$Harvard--Smithsonian Center for Astrophysics (CfA), 60 Garden Street, Cambridge, Massachusetts 02138, USA\\
$^2$Instituto de Astrof\'isica de Canarias (IAC), E--38200 La Laguna, Tenerife, Spain \\
$^3$Department Astrof\'isica, Universidad de La Laguna, E-38206 La Laguna, Tenerife, Spain\\
$^4$Istituto di Astrofisica e Planetologia Spaziali (INAF-IAPS), Via Fosso del Cavaliere 100, I-00133 Roma, Italy\\
$^5$Department Astrof\'isica, CEI Canarias -- Campus Atl\'antico Tricontinental (ULPGC--ULL), Universidad de La Laguna, E-38206, Spain\\
$^6$European Southern Observatory, Alonso de C\'ordova 3107, Vitacura, Casilla 19001, Santiago de Chile, Chile\\
$^7$Max-Planck-Institut f\"ur Astronomie, K\"onigstuhl 17, D-69117 Heidelberg, Germany \\ 
$^8$Finnish Centre for Astronomy with ESO (FINCA), University of Turku, V\"ais\"al\"antie 20, FI-21500 Kaarina, Finland   }

\begin{document}

\date{}

\pagerange{\pageref{firstpage}--\pageref{lastpage}} \pubyear{2007}

\maketitle

\label{firstpage}

\begin{abstract}
The distribution of warm molecular gas (1000--3000 K), traced by the near-IR H$_2$ 2.12 $\mu$m line, has been imaged with a resolution $<0.5$ arcsec in the central 1 kpc of seven nearby Seyfert galaxies. We find that this gas is highly concentrated towards the central 100 pc and that its morphology is often symmetrical. Lanes of warm H$_2$ gas are observed only in three cases (NGC\,1068, NGC\,1386 and Circinus) for which the morphology is much wider and extended than the dust filaments. We conclude that there is no one-to-one correlation between dust and warm gas. This indicates that, if the dust filaments and lanes of warm gas are radial streaming motions of fuelling material, they must represent \textit{two different phases of accretion}: the dust filaments represent a colder phase than the gas close to the nucleus (within $\sim$100 pc). We predict that the morphology of the nuclear dust at these scales should resemble that of the cold molecular gas (e.g. CO at 10--40 K), as we show for CenA and NGC\,1566 by Atacama Large Millimeter/submillimeter Array (ALMA) observations, whereas the inner H$_2$ gas traces a much warmer phase of material identified with warmer (40-500 K) molecular gas such as CO(6-5) or HCN (as shown by ALMA for NGC\,1068 and NGC\,1097).
We also find that X-ray heating is the most likely dominant excitation mechanism of the H$_{2}$ gas for most sources.
\end{abstract}



\section{Introduction}
Understanding how supermassive black holes (SMBHs) grow and how this is connected to the formation and evolution of their host galaxy is one of the major questions in astrophysics (e.g. see review by \citealt{2013ARA&A..51..511K}). The accretion of gas coming from galactic scales ($\sim$10 kpc) is considered to be the main fuelling mechanism of SMBHs and eventually of active galactic nuclei (AGN). In order to trigger and feed nuclear activity, gas must be transported from kpc scales to a few Schwarzschild radii around the SMBH and lose its large angular momentum. 
At kpc-scales, galaxy interactions and mergers and large-scale bars have been proposed as the most efficient mechanisms driving gas into the inner kpc region (e.g. \citealt{2006ApJS..163....1H}; \citealt{2008A&A...492...31D}; see \citealt{2006asup.book..159C} and \citealt{2006LNP...693..143J} for a review). On smaller scales (a few 100 pc), nested bars (\citealt{1989Natur.338...45S}), nuclear spirals (\citealt{2004MNRAS.354..883M}) and radiation pressure driven inflows (\citealt{2012ApJ...758...66W}) are proposed mechanisms for driving matter to the centre.

A nuclear spiral produced in a gaseous medium by shocks induced by the large-scale bar may also drive the gas to the central parsecs of the galaxy (e.g. \citealt{2000ApJ...528..677E}; \citealt{2005AJ....130.1472P}; \citealt{2007ApJ...670..959S}; \citealt{2009ApJ...702..114D}). Dusty spirals are also a common phenomenon in both Seyfert and ``normal'' galaxies (\citealt{2003ApJ...589..774M}), and provide an alternative mechanism for the fuelling process. The presence of nuclear dust filaments connected to kpc-scale dust structures was observed in several Seyfert galaxies (e.g. \citealt{1998ApJS..117...25M}; \citealt{2004ApJ...616..707H}; \citealt{2014MNRAS.442.2145P}), suggesting that these are the channels of inflow of material from the outer parts of the galaxy to the nuclear region. Probing a connection between the dusty filaments and gas distribution at scales of a few 100 pc and AGN feeding has been possible only in a few nearby AGN using high-angular resolution observations of the cold molecular gas (e.g. \citealt{2009A&A...496...85G}; \citealt{2011ApJ...736...37K}; \citealt{2013A&A...558A.124C,2014A&A...565A..97C}) or of warm gas tracers such as the H$_{2}$ molecular gas (e.g. \mbox{\citealt{2006A&A...454..481M}}; \mbox{\citealt{2007ApJ...671.1329N}}; \citealt{2009ApJ...702..114D}; \citealt{2009ApJ...696..448H, 2013ApJ...768..107H}; \citealt{2013MNRAS.428.2389M}). The latter is usually traced by integral-field spectroscopy (IFU) observations limited to a small instrumental field of view (FoV), often of less than 3 arcsec $\times$ 3 arcsec. For instance, evidence of a close relation between dust and molecular gas has been observed in the Seyfert galaxies NGC\,1566 and NGC\,1433, for which recent Atacama Large Millimeter/submillimeter Array (ALMA) observations reveal a nuclear spiral of cold CO(3--2) molecular gas that is fuelling the nucleus and is spatially coincident with the dust structures seen at scales $\leq$ 400 pc (\citealt{2013A&A...558A.124C,2014A&A...565A..97C}). In NGC\,4501, the warm H$_{2}$ molecular gas traced by VLT/SINFONI\footnote{Spectrograph for INtegral Field Observations in the Near Infrared (SINFONI) on the Very Large Telescope (VLT).} within a FoV of 3 arcsec $\times$ 3 arcsec correlates very well with the filamentary dust lanes observed at scales $<$100 pc (\citealt{2013MNRAS.428.2389M}). It should be noted though that kinematic studies of the molecular warm gas provide controversial results concerning the presence of molecular outflows in the central regions of some AGN, as is the case of NGC\,1068, for which the results of the SINFONI observations were interpreted as an inflow of warm H$_{2}$ gas towards the nucleus at scales of a few pc that could be fuelling the AGN (\citealt{2009ApJ...691..749M}) but as an outflow by \cite{2014MNRAS.445.2353B}. Unfortunately, the small FoV of IFU observations is in many cases insufficient to fully characterize the gas kinematics as a large fraction of gas may reside outside the covered IFU FoV, as it is shown in this work. 

In this paper we show the full extension of the warm ($\sim$1000-3000 K) H$_{2}$ gas of a nearby sample of five type 2 Seyfert galaxies (CenA, Circinus, NGC\,1068, NGC\,1386 and NGC\,7582), one type 1.5 Seyfert (NGC\,1566) and one type 1 LINER/Seyfert (NGC\,1097), finding that the warm molecular gas is often detached from the central dust lanes and filaments that are observed at equivalent angular scales as the H$_{2}$ gas ($\sim$ 0.1--0.4 arcsec). This may be pointing to two modes of accretion: a cold mode, traced by the dust filaments, and a warm mode, traced by the warm H$_{2}$ gas and warm molecules such as HCN or CO of high $J$ levels traced by ALMA. The existence of two distinct accretion modes, hot ($10^{5-6}$ K) and cold, has been shown by observations (e.g. \citealt{2013ApJ...777..163H}), predicted by high-resolution simulations of black hole accretion on sub-kpc scales (e.g. \citealt{2013MNRAS.432.3401G}; \citealt{2014MNRAS.441.3055B}) and implemented in cosmological simulations by \cite{2015MNRAS.448.1504S} in an attempt to better describe a smooth transition between the so-called radio- and quasar-modes of accretion. The existence of an intermediate warm gas mode of accretion should therefore take place. We tackle this aspect in this paper by investigating the association between relatively warm ($\sim$1000-3000 K) H$_{2}$ gas and cold dust filaments seen in detail in the central few 100 pc of active galaxies. The targets have been drawn from a larger sample that includes the nearest and brightest AGN that could be observed at high resolution in the infrared (IR) using the VLT and adaptive optics (AO; \citealt{2010MNRAS.402..724P}; \citealt{2010MNRAS.402..879R}). We present near-IR narrow-band H$_2$ imaging observations with the VLT of this sample of seven nearby AGN. The large FoV of the observations (up to 26 arcsec $\times$ 26 arcsec) and high spatial resolution (down to 0.1 arcsec) allows
  us to study the distribution of the molecular H$_2$ warm gas and its potential as a tracer of dust at scales from a few tens of pc to a few 100 pc as well as to estimate the amount of gas available for accretion on to the SMBH. In addition to the narrow-band imaging data, we present near-IR broad-band imaging data taken with the same instrument, optical \textit{I}- and \textit{V}-band data from the \textit{Hubble Space Telescope} (\textit{HST}), AO VLT/SINFONI data, X-ray imaging data from the \textit{Chandra} X-ray satellite, and CO and HCN intensity maps from ALMA.

The paper is organized as follows. The observations and data reduction are presented in Section~\ref{observations}. The data analysis is described in Section~\ref{analysis}. The results obtained are presented and discussed in Section~\ref{results}, while final conclusions are summarized in Section~\ref{conclusions}. An individual description of each target galaxy is provided in Appendix\,\ref{individual}.


\section{Observations and data reduction}
\label{observations}
\subsection{Adaptive optics IR data}
The seven galaxies listed in Table~\ref{table1} were observed with the near-IR AO
assisted instrument Naos-Conica (NaCo; \citealt{2003SPIE.4839..140R}) installed
at the Nasmyth focus on ESO VLT (Paranal, Chile). Near-IR images in several broad- and narrow-band filters in the \textit{K}-band were taken with NaCo using the pixel scale of 0.027 arcsec pixel$^{-1}$. A FoV of 26 arcsec $\times$ 26 arcsec of the central few 100 pc was covered by NaCo for all the galaxies. In all cases the AO correction was done on the nucleus or a star cluster close to the nucleus using wavefront sensing in the optical. 
For the purpose of this paper, we make use of the $2\, \rm{\micron}$ broad-band \textit{Ks} filter ($\lambda_c = 2.180\, \rm{\micron}$, $\Delta\lambda = 0.350\, \rm{\micron}$) to obtain the continuum image. This should provide the maximum contrast between the nuclear emission and the host galaxy (e.g. \citealt{2010MNRAS.402..724P,2014MNRAS.442.2145P}). 

For all the targets, narrow-band images were taken with two filters, each with different width but both centred on the H$_2$ 1--0 S(1) $2.12\, \rm{\micron}$ line. The filters used were NB2.12 (width $0.022\, \rm{\micron} $) and IB2.12 ($0.06\, \rm{\micron}$). Exposure time in both filters was the same for each galaxy, ranging from 4 to 5 min on-source. Since the seven target galaxies are nearby and have small systemic velocities (see Table~\ref{table1}), the H$_2$ 1--0 S(1) $2.12\, \rm{\micron}$ line is fully covered by the NB2.12 filter. The resolution in these narrow-band images was estimated from point-like sources detected in the FoV. In NGC\,1566 foreground stars were used for resolution estimate, while for the rest the estimate was based on stellar clusters, which are unresolved with 8-m telescopes in the near-IR. For each galaxy, the spatial resolution achieved in the two filters is comparable and ranges from full width at half maximum (FWHM)$\sim$0.12 arcsec for NGC\,1566 to FWHM$\sim$0.35 arcsec for NGC\,1097 (see Table~\ref{table1}).


For the galaxies Cen\,A, Circinus and NGC\,1386, the intermediate-band line-free filter IB2.06 ($\lambda_c = 2.060\, \rm{\micron}$, $\Delta\lambda = 0.060\, \rm{\micron}$) was also used to estimate the continuum level.

The NaCo data reduction was performed following the standard procedure of sky-subtraction, flat fielding and combination of frames with the ESO \textsc{eclipse} package \citep{1999ASPC..172..333D}.

The imaging data are complemented with IFU VLT/SINFONI data for NGC\,1386, NGC\,1566 and NGC\,7582. The data reduction was performed using the SINFONI pipeline provided by \textsc{eso} and includes: correction for bad pixels, flat-field, geometric distortions, wavelength calibration, reconstruction of the data cube from the image spectral slices, background subtraction using sky frames, and flux calibration using a standard star.


\begin{table*}
\begin{minipage}{\textwidth}
\centering
\scriptsize
\caption{Galaxy properties, filters and errors.}\label{table1}
\begin{tabular}{lcccccccccccc}
\hline
Object &       Galaxy   &  AGN  &  $L_\mathrm{bol}$ & $D_{L}$ & Linear   & Ref.     & Dust map & H$_{2}$ gas  & \multicolumn{2}{c}{FWHM}   & Clusters  	& Errors             \\
           &        type	   &	type  &  				& 	    	 &         scale                &            &   filters     &    filters         &   \textit{Ks}     &  NB2.12     & registration    & registration   \\
           &       		   &		  &  [erg s$^{-1}$]  & [Mpc]    	 &    [pc arcsec$^{-1}$]                     &            &   		 &    		        &   		          &       		&   			&  [mas]             \\

(1)       &         (2)       &   (3)    &    		(4)        &        (5)      &    (6)    	    &   (7)       &         (8)    &   (9)       	 &  (10)      	    & (11) 		&       (12)     	& (13)     \\ 
\hline
CenA         &  S0	   &  Sy2   &  6 $\times$ 10$^{42}$  & 3.8       &        17             &  [1]      &  K/F814W & NB2.12-IB2.12 & 0\farcs12 & 0\farcs13  & 4$^{*}$  & 60, 120, 60, 70 \\ 
Circinus     &  SAb      &  Sy2	 & 8.4 $\times$ 10$^{42}$   & 4.2       &        19  	       &   [2]     &  K/F814W & NB2.12-IB2.12 & 0\farcs16 & 0\farcs16  &     3        & 40, 30, 30         \\ 
NGC\,1068 &  Sb        &  Sy2   &  8.7 $\times$ 10$^{43}$  & 14.1      &        70            &   [3]     &  K/F550M & NB2.12-IB2.12 & 0\farcs14 & 0\farcs13  &     3 	       & 30, 30, 10         \\
NGC\,1097 &  SBb	   & L1/Sy1&  1.5 $\times$ 10$^{41}$   & 14.2	            &        69  	       &   [4]     &  K/F814W & NB2.12-IB2.12 & 0\farcs14 & 0\farcs35  &    7         & 		20,30,30,30,20,30,30           \\
NGC\,1386 &  SBa	   &  Sy2	 &  1.2 $\times$ 10$^{42}$  &  15.3       &       74  	       &   [5]     &  K/F814W & NB2.12-IB2.12 & 0\farcs09 & 0\farcs19  &     2$^{*}$          &    10, 10                     \\
NGC\,1566 &  SABbc  & Sy1.5	 &  5.5 $\times$ 10$^{42}$  &  20.5       &       99  	       &   [6]     &  K/F438W & NB2.12-IB2.12 & 0\farcs13 & 0\farcs12  &     3        &    40, 40, 10        \\
NGC\,7582 & SBab	   & Sy2    & 1.8 $\times$ 10$^{43}$   & 19.9      &        96           &   [7]     &  F160W/F606W & SINFONI$^{**}$ & 0\farcs16 & 0\farcs24  &  4	&	20, 50, 70, 20	     \\
\hline
\end{tabular}
\end{minipage}
\raggedright
\smallskip\newline\small {\bf Column designation:}~(1) Object name; (2) galaxy type; (3) AGN type: Sy stands for Seyfert, L for LINER; (4) bolometric luminosity, from \cite{2014ApJ...787...62M} and \cite{2010MNRAS.402..724P}; (5) luminosity distance; (6) linear scale; (7) references for column (5): [1] \cite{2010PASA...27..457H}; [2] \cite{1977A&A....55..445F}; [3] \cite{1997A&A...320..399M}; [4] \cite{2009AJ....138..323T}; [5] \cite{2003ApJ...583..712J}; [6] NED adopting $H_0$ = 73 {\kms} Mpc$^{-1}$; [7] \cite{2002A&A...393...57T}; (8) and (9) filters used to construct the dust maps and H$_{2}$ gas maps, respectively; (10) and (11) FWHM of the \textit{Ks}-band and NB2.12 filters; (12) number of stars/clusters used in the alignment; (13) error on the position of each star/cluster used in the alignment.\\
$^*$ for CenA and NGC\,1386, the NB2.12 and IB2.12 images have been aligned to the NaCo/\textit{K}-band image using the nucleus; $^{**}$ for NGC\,7582 no good quality map with NaCo is available. The H$_{2}$ molecular gas map of this galaxy is thus derived from SINFONI data.
\end{table*}

\subsection{\textit{HST} optical data}
To construct the colour dust maps, we searched in the \textit{HST} archive for the reddest image available for each galaxy. 
This should provide the largest number of common sources to the 2$\, \rm{\micron}$ NaCo images to use as reference sources in the image registration. 
Broad \textit{I}-band images corresponding to the \textit{HST/F814W} filter from either the WFPC2\footnote{Wide-Field Planetary Camera 2.} or the WFC3\footnote{Wide-Field Camera 3.} cameras were retrieved for most of the target galaxies (see Table~\ref{table1}). When not available, the \textit{V}-band image from the ACS/\textit{F550M} filter was used. In the case of NGC\,7582, the colour maps were built using the WFPC2/\textit{F606W} and NICMOS\footnote{Near Infrared Camera and Multi-Object Spectrometer.}/\textit{F160W} images to avoid the strong contribution of the nucleus in the \textit{K}-band. In all cases the pixel scale ranges from 0.04 to 0.05 arcsec pixel$^{-1}$.

For data sets with multiple exposures, the retrieved images were combined and corrected for geometric distortions using the \textit{MultiDrizzle} package in PyRAF (\citealt{2003hstc.conf..337K}). Cosmic rays removal was applied for those datasets with single exposures using the L.A. Cosmic algorithm (\citealt{2001PASP..113.1420V}) in its \textit{Python} module version\footnote{\url{http://python.org}\\ \url{http://obswww.unige.ch/~tewes/cosmics_dot_py/}}. Further analysis was performed with \textsc{iraf}/PyRAF.

\section{Analysis}
\label{analysis}
\subsection{Extraction of H$_2$-line maps}
\label{h2maps}
The extraction of the continuum-free H$_2$ images was performed in two steps. First, the subtraction of the two images centred on the $2.12\,\rm{\micron}$ H$_2$ line (IB2.12--NB2.12) produced a pure continuum (offline) image. This pure-continuum image was further used to extract pure H$_2$ images from the narrow-band NB2.12 image using a scaling factor. This process was repeated iteratively, changing the scaling factor, until the residuals of the H$_2$ emission-line image were smooth. Effectively this corresponds to $H_{2} = NB - c\times(IB-NB)$, where $c$ is the scaling factor. This process was performed to obtain the H$_2$ maps of all the galaxies in the sample except for NGC\,7582 (see below).

For Cen\,A, Circinus and NGC\,1386, H$_2$ maps were also extracted using the filters IB2.12 and IB2.06, containing the emission line and adjacent continuum, respectively. A similar iterative process was applied to extract the H$_2$ emission-line maps. The latter were compared to those based on the NB2.12 filter. In all cases, similar H$_{2}$ structures are derived using these two approaches.

In addition to the NaCo narrow-band imaging, we show for all sources (Figs.~\ref{cena}-\ref{n7582}) the molecular H$_{2}$ gas distribution obtained from AO VLT/SINFONI IFU observations. For Cen\,A, NGC\,1068, NGC\,1097, NGC\,1386 and NGC\,1566, the H2 line maps have been extracted from the SINFONI data cube using \textsc{qfitsview}\footnote{\url{http://www.mpe.mpg.de/~ott/QFitsView/}}. The 3D data cube allows an optimized extraction of the line emission, subtracting the adjacent continuum measured at both sides of the line. The warm SINFONI H$_{2}$ maps for Circinus and NGC\,1068 are directly from \cite{2006A&A...454..481M} and \cite{2009ApJ...691..749M}, respectively.

For NGC\,7582, the IB2.06 filter could not be used as reference for the continuum level due to contamination from the \textsc{He\,i} emission line at $2.058\, \rm{\micron}$, associated with the circumnuclear starburst. Thus, the H$_2$ line map shown in Fig.\,\ref{n7582} was extracted from the SINFONI data cube following the same procedure as for CenA, NGC\,1097, NGC\,1386 and NGC\,1566.

\subsection{Image registration}
For each target galaxy, the \textit{HST} images were registered to the NaCo \textit{K}-band image and the H$_2$ narrow-band
images using at least two point-like reference sources (usually unresolved stellar clusters in the galaxy and avoiding the nucleus when possible) in the common FoV to all images. In general, up to three to four reference sources could be identified (see Table~\ref{table1}) due to the limited FoV and the strong galaxy emission in the near-IR.
The image registration was performed with the IRAF task \textit{imalign} and crosschecked with an alignment script written by us in IDL\footnote{Interactive Data Language.} following the same procedure as in \cite{2014MNRAS.442.2145P}.
The relative offsets between the images were determined using the centroid position of the reference sources identified in the common FoV and never using the nucleus to register the near-IR and optical images (see \mbox{\citealt{2014MNRAS.442.2145P}}). The near-IR nucleus was only used to align the \textit{K}-band image to the H$_2$ narrow-band image in the case of CenA, NGC\,1386 and SINFONI for NGC\,7582. The average of these offsets was used to shift and align the images. For each target, the alignment error of each reference source is estimated as the standard deviation of its position in the different images (Table~\ref{table1}, column 13). The mean of these individual errors is taken as the final error of the alignment for each target galaxy, which ranges between 20 and 80 mas. Therefore, any uncorrected distortions or wavelength-dependent effects affecting the centroid position of the point-like reference sources used in the image registration are accounted for in the final alignment error. See \cite{2014MNRAS.442.2145P} for further details.

\begin{figure}
 \includegraphics[width=0.5\textwidth]{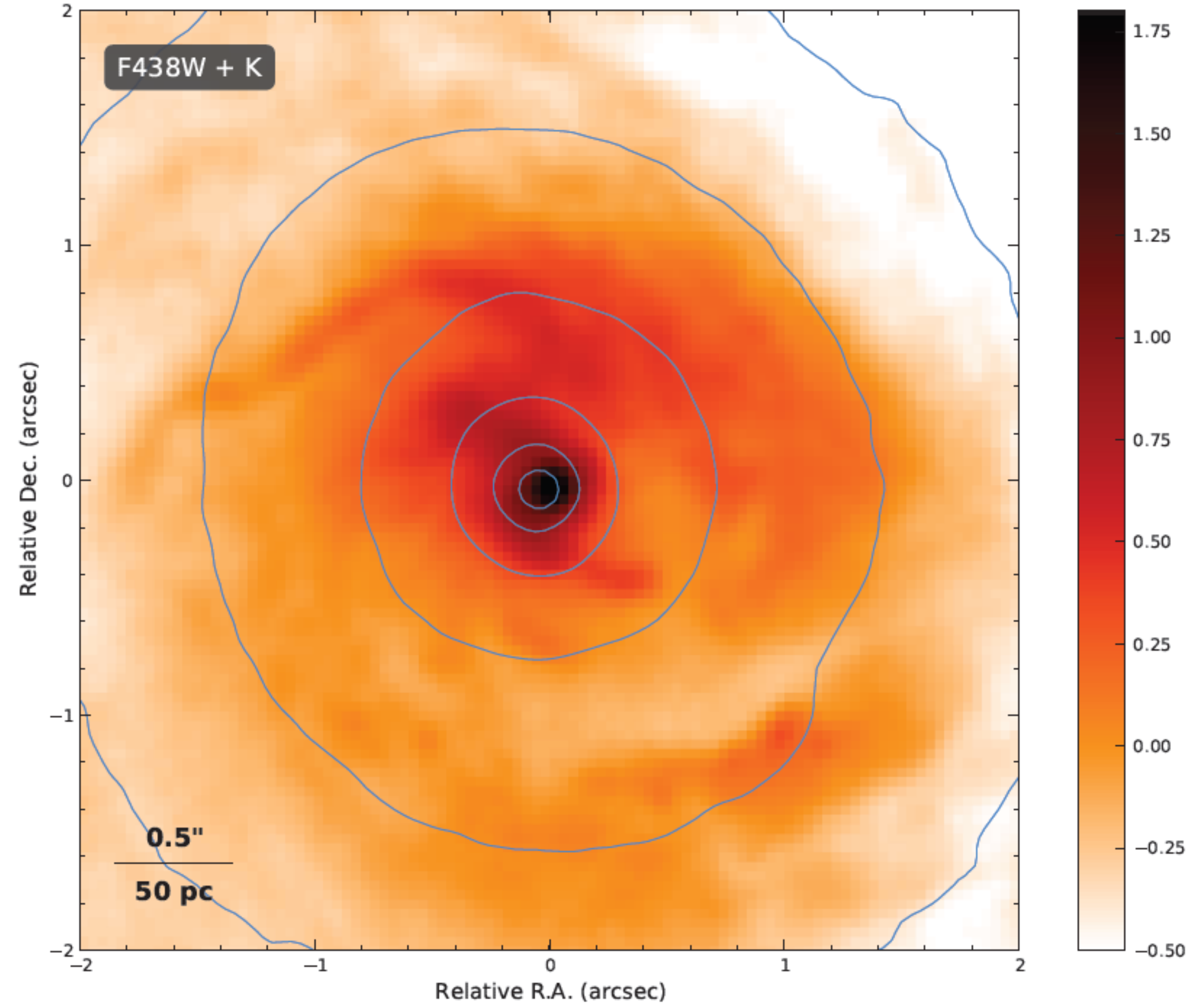}
 \protect\caption[n1566]{NGC\,1566. \textit{HST/F438W} image with NaCo/\textit{K}-band continuum contours in blue. The FoV is 4 arcsec $\times$ 4 arcsec. The colour scale is in arbitrary units. North is up and east is to the left.}
 \label{n1566shift}
\end{figure}

The accurate registration of the \textit{Ks}-band and \textit{HST} optical images allows us to confirm the shift between the \textit{HST/F814W} and NaCo/\textit{Ks}-band peaks previously reported for the type 2 Seyfert Circinus and to constrain it to $160 \pm 30\, \rm{mas}$ (Mezcua et al., in preparation). For the same galaxy we also note a shift of $40 \pm 30\, \rm{mas}$ between the peak of the warm H$_2$ molecular gas and the \textit{Ks}-band peak. In addition, we report for the first time an offset between the optical (\textit{HST/F438W}) and NaCo/\textit{Ks}-band peaks of emission for the type 1.5 Seyfert NGC\,1566 of $50 \pm 30\, \rm{mas}$ (Fig.~\ref{n1566shift}). 

The NaCo warm molecular H$_2$ gas emission map of Circinus, NGC\,1097, NGC\,1386 and NGC\,7581 has been also aligned to their archival broad-band X-ray \textit{Chandra} image (bottom right panel in Figs.~\ref{circinus}, \ref{n1097}, \ref{n1386}, and \ref{n7582}). The image registration has been performed by matching the position of the H$_2$ peak to that of the nuclear X-ray peak due to the lack of point-like references sources in the common FoV of the H$_2$ and \textit{Chandra} images. In the case of NGC\,1068, the SINFONI warm H$_2$ gas map has been aligned to the \textit{Chandra} High Resolution Camera (HRC) image of the nuclear region of NGC\,1068 (\citealt{2012ApJ...756..180W}) by matching the position of the nucleus in the \textit{Ks}-band to the nuclear X-ray peak (Fig.~\ref{n1068}, middle right panel). This same procedure (i.e. matching the position of the nucleus) has been performed to align the SINFONI warm H$_2$ gas map of NGC\,1068 to its ALMA CO(6-5) map (\citealt{2014A&A...567A.125G}) and the NaCo warm molecular H$_2$ gas emission image of NGC\,1097, NGC\,1566 and CenA to their ALMA HCN  (\citealt{2013PASJ...65..100I}; \citealt{2015A&A...573A.116M}), CO(3-2) (\citealt{2014A&A...565A..97C}) and CO(2-1) maps (\citealt{2013ASPC..476...69E}), respectively.

\subsection{Dust maps}
 \textit{I - K} or \textit{V - K} colour maps of the central few $100\, \rm{pc}$ of each target galaxy were constructed using the ratio between the aligned NaCo \textit{K}-band and \textit{HST} optical images (Figs.~\ref{cena}--\ref{n7582}). These maps show the presence of dust in the central few $100\, \rm{pc}$ of each galaxy. The dust morphology is in general filamentary, in some cases more disc-like or with irregular shape. In Cen\,A, Circinus, NGC\,1068, NGC\,1386 and NGC\,7582 (i.e. those type 2 Seyferts) a filament of dust is observed to cross the nucleus, unequivocally identified as an outstanding point-like source in the NaCo $2\, \rm{\micron}$ \textit{Ks}-band images (e.g. \citealt{2010MNRAS.402..724P,2014MNRAS.442.2145P}; \citealt{2012JPhCS.372a2006F}). Such a nuclear dust lane is not observed in NGC\,1566 (Sy1.5) nor NGC\,1097 (Sy1). In NGC\,1097 the dust is observed to spiral around the nucleus and to recover the same morphology, although with much larger contrast, as that originally discovered in the high-angular resolution \textit{J}-\textit{K} maps of \cite{2005AJ....130.1472P}. In some sources (e.g. NGC\,1386) the central dust filaments extend to $\sim 1\, \rm{kpc}$ and are thus associated with the large-scale dust structure of the galaxies (see also \citealt{2014MNRAS.442.2145P}).

For CenA, NGC\,1068, NGC\,1386 and NGC\,1566 the warm molecular H$_2$ gas emission map from SINFONI has been aligned to the dust map by matching the position of the H$_2$ peak to that of the nucleus (or NaCo/$K$-band peak; e.g. \citealt{2009ApJ...691..749M}). All maps are centred at the \textit{K}-band peak and are shown in Figs. ~\ref{cena}--\ref{n7582}.

\subsection{Molecular H$_2$ warm gas light profile}
\begin{figure}
 \includegraphics[width=0.48\textwidth]{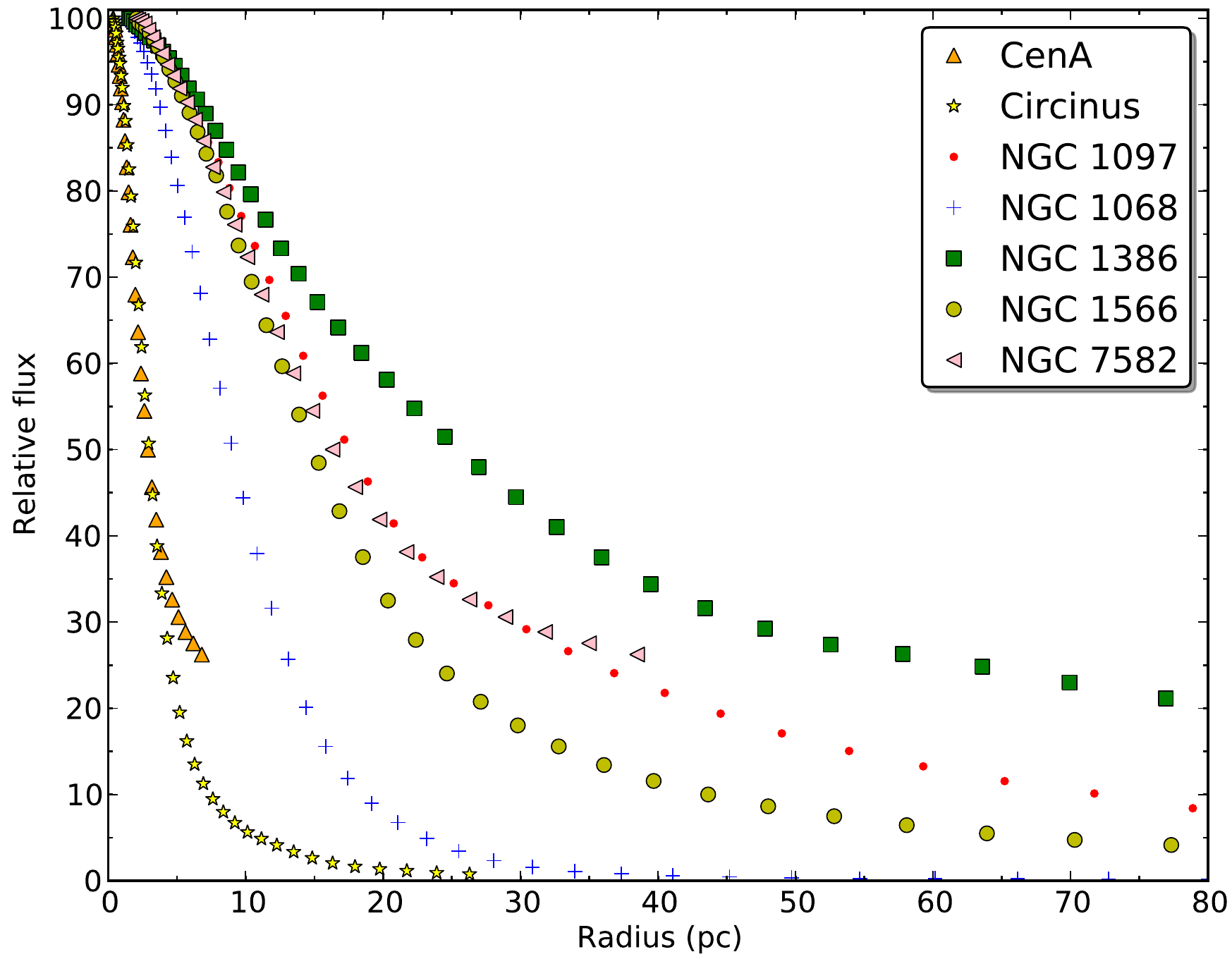}
 \protect\caption[perfil]{Distribution of the warm H$_2$ gas provided by the fit of circular isophotes (elliptical for NGC\,1068 and NGC\,1386) to the NaCo emission (SINFONI for NGC\,7582). Fitting a S\'ersic function indicates an exponential profile for all sources, with a mean HWHM of 27 pc.}
 \label{perfil}
\end{figure}

To quantify the extent of the nuclear H$_2$ warm gas distribution traced by NaCo, circular or elliptical isophotes were fitted to the central regions of the H$_2$ emission-line images (i.e. we exclude the H$_2$ clumps associated with star formation observed in the spiral arms of NGC\,1097 or the extended starburst of NGC\,7582). The extent of these isophotes, taken as three times the background noise level ($\sigma$) of each H$_2$ emission-line image, is reported in Table~\ref{table2} (see also Fig.~\ref{perfil}). A S\'ersic function (\citealt{1968adga.book.....S}) was fitted to the light profiles of the warm H$_2$ gas shown Fig.~\ref{perfil}.


\begin{table*}
\begin{minipage}{\textwidth}
\centering
\caption{Properties of the molecular H$_{2}$ gas.}\label{table2}
\begin{tabular}{lcccccccc}
\hline
Object &       \multicolumn{2}{c}{Radius}    	&	$L_\mathrm{H2}$ 	&    HWHM	&$M_\mathrm{warm}$ 	& $M_\mathrm{cold}$ & $N_\mathrm{cold}$	& $L_\mathrm{X-rays}$	 \\
           &      [arcsec]  		& [pc] 		&	[erg s$^{-1}$]	        &    	[pc]		&	[M$_{\odot}$]	  	&  [M$_{\odot}$]    	  & [cm$^{-3}$]			&	[erg s$^{-1}$]		 \\
(1)       &         (2)       		&   (3)    		&            (4)        	  	&       (5)   	&		(6) 			&        (7)             		  &	(8)			&		(9)			 \\ 
\hline
CenA         		&    2.0    	&	34		&	$2.4\times10^{37}$	&      3		&      	50		 	&  $7\times10^{6}$	 &	$9\times10^{3}$	&	$1.1\times10^{42}$	\\ 
Circinus$^{*}$     	&    1.4    	&	26		&	$3.9\times10^{37}$	&      4		&      	40		  	&  $1\times10^{7}$ 	 &	$3\times10^{4}$	&	$4.3\times10^{41}$	\\
NGC\,1068$^{\dagger}$&  1.5	&	106		&	$1.7\times10^{39}$	&	10		&		1000			&  $5\times10^{8}$	 &	$6\times10^{4}$	&	$6.5\times10^{41}$	\\
NGC\,1097$^{*}$ 	& 	2.0	&	138		&	$1.3\times10^{39}$	&	29		&		600			&  $4\times10^{8}$	 &	$6\times10^{4}$	&	$5.2\times10^{40}$	\\
NGC\,1386$^{\dagger}$& 	4.3	&	321	        &	$1.8\times10^{39}$	&	100		&		1000  		&  $6\times10^{8}$	 &	$5\times10^{4}$	&	$1.3\times10^{40}$	\\
NGC\,1566 		&     2.2    &	220		&	$4.5\times10^{39}$	&      27		&	       3000		  	&  $1\times10^{9}$	 &	$6\times10^{3}$	&	$3.7\times10^{42}$	\\
NGC\,7582$^{*,\dagger\dagger}$& 0.3&29		&	$1.6\times10^{38}$	&	14		&		100			&  $5\times10^{7}$	 &	$1\times10^{5}$	&	$4.5\times10^{42}$	\\    
\hline
\end{tabular}
\end{minipage}
\raggedright
\smallskip\newline\small {\bf Column designation:}~(1) Object name; (2, 3) aperture radius used to extract the flux of the warm H$_{2}$ molecular gas, in arcsec and pc respectively; (4) luminosity of the warm H$_{2}$ molecular gas; (5) HWHM of the S\'ersic funcion fit to the warm H$_{2}$ gas distribution; (6) mass of the warm H$_{2}$ molecular gas; (7) mass of the total (cold) H$_{2}$ molecular gas; (8) density of the total (cold) H$_{2}$ molecular gas assuming a gas-to-mass ratio $f_\mathrm{g}$ = 1\% for NGC\,1097 and $f_\mathrm{g}$ = 10\% for the rest; (9) X-ray luminosity from the highest energy range available in the literature for CenA, Circinus, NGC\,1068, NGC\,1097, NGC\,1566, and NGC\,7582 (\citealt{2010MNRAS.402..724P} and references therein). For NGC\,1386, the X-ray luminosity has been derived from the 2-10 keV X-ray flux from \cite{1997PASJ...49..425I} assuming a luminosity distance of 15.3 Mpc. \\
$^*$ for Circinus, NGC\,1097 and NGC\,7582, in addition to the AGN-associated warm gas, there are clumps of H$_{2}$ gas associated with star formation; $^{\dagger}$ For NGC\,1068 and NGC\,1386, elliptical apertures of ellipticity 0.6 and 0.7 were used, respectively; $^{\dagger\dagger}$ The H$_{2}$ molecular gas properties of NGC\,7582 come from SINFONI data.
\end{table*}

\section{Results and discussion}
\label{results}
\subsection{Morphology of the molecular H$_2$ warm gas}
\label{morphology}
Molecular H$_2$ warm gas emission is seen in the central regions of all the galaxies discussed in this paper (see Figs.~\ref{cena}--\ref{n7582}). The emission is resolved by NaCo in all cases, and its morphology is varied, ranging from symmetric emission to showing extended filaments. In some cases, the warm H$_2$ gas is also observed as clumps of emission in the outermost spiral arms (e.g. NGC\,1097, see Fig.~\ref{n1097}). The gas is thus observed when it is close enough to the nucleus as to be heated by the AGN, though it is equally detected when heated by star-forming regions (also visible in our NaCo \textit{K}-band images). 
A circular aperture was able to reproduce the morphology of the molecular H$_2$ warm gas for all sources except for NGC\,1386 and NGC\,1068, for which an elliptical aperture was found to better reproduce the observed morphology.
In the case of Circinus, additional clumpy H$_2$ warm gas emission is observed south and north of the nucleus at a 3$\sigma$ level. The northern clumps seem to be associated with star formation, while the southern ones are most likely heated by the nuclear source (see Appendix\,\ref{individual}). Taking these clumps into account, we estimate for Circinus a total extent of the nuclear H$_2$ warm gas of 2 arcsec (38 pc) and an average extent of the nuclear H$_2$ warm gas emission for the whole sample of 127 pc. This total extent was defined as the maximum size of the 3$\sigma$ distribution in our maps. The fit of a S\'ersic function to the light profiles of the warm H$_2$ gas (Fig.~\ref{perfil}) indicates exponential distributions (\textit{n}$\sim$1) for all sources and that the warm H$_2$ gas emission is concentrated at the centre with an average half width at half-maximum (HWHM) of 27 pc. 
This HWHM is in agreement with the average extent of the H$_2$ warm gas emission in Seyfert galaxies when traced by IFU observations (e.g. \citealt{2009ApJ...696..448H} find an average HWHM$\sim$30 pc for a sample of Seyfert galaxies that includes Circinus, NGC\,1068 and NGC\,1097 observed by SINFONI with a FoV of up to 3.2 arcsec $\times$ 3.2 arcsec).
When using a larger FoV (e.g. up to 8 arcsec $\times$ 8 arcsec; \citealt{2013MNRAS.428.2389M}; \citealt{2013ApJ...768..107H}), the molecular H$_2$ warm gas is observed to extend up to a few 100 pc (e.g. 144 pc for NGC\,4536; \citealt{2013MNRAS.428.2389M}; up to 250 pc for the five Seyfert galaxies of \citealt{2013ApJ...768..107H}) similarly to what we find for our sample when observed by NaCo. \cite{2013ApJ...768..107H} also study the extent and luminosity of a matched-sample of quiescent galaxies, finding that these present a flatter luminosity profile and lower H$_2$ luminosities within a radius of $\sim$100 pc than their sample of Seyferts. The sample of seven Seyfert galaxies we study in this paper has an average nuclear H$_2$ luminosity of $1.4\times10^{39}$ erg s$^{-1}$ ($3.6\times10^{5}$ L$_{\odot}$; see Table~\ref{table2}), which is one order of magnitude higher than that of the sample of quiescent galaxies of \cite{2013ApJ...768..107H} and of the only non-AGN (NGC\,3351) included in the sample of \cite{2013MNRAS.428.2389M}, and a more centrally concentrated nuclear H$_2$ distribution than that of the quiescent/non-AGN. These results are in agreement with those found by \cite{2013ApJ...768..107H}, suggesting that the nuclear H$_2$ molecular gas in the AGN constitutes a gas reservoir of material available for fuelling the central SMBH (see also Section~\ref{mass}).

\subsection{NaCo versus SINFONI}
The large FoV of the narrow-band NaCo imaging allows us to trace the distribution of the warm molecular H$_{2}$ gas beyond the instrument-limited small nuclear regions traced by IFU instruments as SINFONI. The different extent and morphology of the molecular H$_2$ warm gas traced by NaCo compared to that traced by a SINFONI FoV of up to $\sim$ 2 arcsec $\times$ 2 arcsec can be directly observed in Figs.~\ref{cena}--\ref{n1566} for CenA, Circinus, NGC\,1097, NGC\,1068, NGC\,1386 and NGC\,1566. In all these cases the H$_2$ warm gas in the NaCo maps extends beyond the SINFONI FoV, up to $\sim$ 6 -- 8 arcsec (a few $100\, \rm{pc}$) for most galaxies. For some sources like Circinus, the gas presents a complex morphology very distinct to the relatively symmetric distribution traced by SINFONI, while in others like NGC\,1097 the extension of the H$_{2}$ emission seems to have a different orientation for NaCo and SINFONI. These morphological differences are not caused by a higher SINFONI spatial resolution, as this is in many cases comparable or even lower than that of NaCo (e.g. for NGC\,1097 the SINFONI spatial resolution is FWHM$_\mathrm{SINFONI}$=0.25 arcsec, \citealt{2009ApJ...702..114D}, while FWHM$_\mathrm{NaCo}$=0.14 arcsec, see Table~\ref{table1}; for CenA FWHM$_\mathrm{SINFONI}$=0.12 arcsec, \citealt{2007ApJ...671.1329N}, while FWHM$_\mathrm{NaCo}$=0.12 arcsec; for Circinus FWHM$_\mathrm{SINFONI}$=0.22 arcsec, \citealt{2006A&A...454..481M}, while FWHM$_\mathrm{NaCo}$=0.16 arcsec). The differences between the NaCo and SINFONI H$_2$ intensity maps can be ascribed to the technique used to extract the H$_2$ line map from the SINFONI data cube: the continuum subtraction from the SINFONI data cube (e.g. in the case of NGC\,1068; \citealt{2009ApJ...691..749M}) relies on the equivalent width of the CO bands. Thus, the derived H$_2$ emission and morphology are quite uncertain in the nucleus, where the CO bands are almost diluted by the contribution of the non-stellar emission. This can seriously affect the morphology of the inner regions, as clearly evidenced for NGC\,1068: a nuclear hole is observed in the warm H$_2$ emission-line map of \cite{2009ApJ...691..749M} (see SINFONI map in Fig.~\ref{n1068}), which is not present in the NaCo H$_2$ emission map. We note that such nuclear hole is not present either in the warm H$_2$ emission map of \cite{2014MNRAS.445.2353B}, in agreement with the NaCo map presented in this work. Therefore, the NaCo image does recover more H$_2$ emission than the SINFONI maps shown in Fig.~\ref{n1068}. 

In order to check for consistency, we extract the H$_2$ 1--0 S(1) $2.12\, \rm{\micron}$ line flux within an aperture in the NaCo H$_{2}$ image equivalent to that of the SINFONI FoV (see Table~\ref{table3}). This is performed for those galaxies for which both NaCo and SINFONI molecular H$_{2}$ gas fluxes are available (CenA, Circinus, NGC\,1097, NGC\,1386, and NGC\,1566). For NGC\,1068 such a comparison is not viable due to the nuclear hole in the SINFONI H$_2$ emission-line map. 
Given the higher sensitivity of NaCo, the H$_2$ 1--0 S(1) line fluxes measured with this instrument are expected to be higher than those obtained with SINFONI. This is indeed the case for Circinus, NGC\,1097 and NGC\,1386, for which the NaCo H$_2$ 1--0 S(1) line flux is one to three orders of magnitudes larger than that from SINFONI. However, for CenA and NGC\,1566 the NaCo H$_2$ 1--0 S(1) line flux is one to two orders of magnitude smaller than the SINFONI flux. No polluting lines or features are observed in the region of the SINFONI spectrum close to the H$_2$ 1--0 S(1) line, indicating that the H$_2$ 1--0 S(1) line flux difference must be caused either by the flux calibration of the SINFONI data cubes or the continuum subtraction in the NaCo narrow-band filter. Calibration problems in the SINFONI data cube for the particular case of CenA could explain flux differences between NaCo and SINFONI of up to a factor 4, since the SINFONI observations were acquired under variable conditions (Burtscher, private communication). For instance, \cite{1990A&A...227..342I} determined a well-calibrated H$_2$ 1--0 S(1) flux in a 6 arcsec aperture of (4.2 $\pm$ 0.5) $\times$ 10$^{-14}$ erg cm$^{-2}$ s$^{-1}$, which is of the same order as the NaCo flux.

\begin{table}
\centering
\caption{Comparison of NaCo and SINFONI fluxes}
\label{table3}
\begin{tabular}{lccc}
\hline
\hline 
			&					&	\multicolumn{2}{c}{H$_2$ 1--0 S(1) flux} 				 \\
Object	     	&      Aperture  	  		&     NaCo   & SINFONI  								 \\
	       		&      [$\arcsec$]       	&   [erg cm$^{-2}$ s$^{-1}$] &[erg cm$^{-2}$ s$^{-1}$]		\\
\hline
CenA      		&	      3$\times$3  	& $1.5\times10^{-14}$     	&  $3.4\times10^{-13}$   		 	\\	
Circinus  		&	   0.8$\times$0.8  	&   $1.2\times10^{-14}$    &  $9.0\times10^{-15}$ $^{*}$   		\\	
NGC\,1097	& 		4$\times$4	&   $5.9\times10^{-14}$ 	&  $9.1\times10^{-15}$			\\
NGC\,1386	& 		4$\times$4	&   $7.3\times10^{-14}$	&  $2.2\times10^{-14}$			\\
NGC\,1566	&	    3$\times$3   	&  $7.7\times10^{-14}$     &  $5.8\times10^{-13}$      		\\
\hline
\end{tabular}
\raggedright
\smallskip\newline\small 
$^*$ The SINFONI H$_2$ 1--0 S(1) line flux of Circinus is taken from \cite{2006A&A...454..481M}.
\end{table}

\subsection{Comparison with dust and cold molecular gas}
To characterize whether the molecular H$_{2}$ warm gas follows the dust and/or is coincident with regions of star formation, the H$_2$ warm gas contours are plotted on top of the dust maps of each galaxy and on top of the \textit{K}-band or \textit{I}-band maps for those galaxies with clumps of star formation. An individual description of each galaxy and of its H$_2$ and dust morphology is provided in the Appendix\,\ref{individual}.
In CenA, NGC\,1097, NGC\,1566 and NGC\,7582 the nuclear warm H$_{2}$ gas presents a diffuse, symmetrical morphology that does not correlate with that of the dust. The observed differences are not caused by a different FWHM between the dust maps and H$_{2}$ (NB2.12) images (see Table~\ref{table1}), but reflect a real different morphology. In Circinus (southern emission), NGC\,1068 and NGC\,1386 the morphology of the H$_{2}$ gas seems to reproduce that of the dust filaments close to the nucleus. However, the correspondence between the warm H$_{2}$ gas and the dust is not one to one. The lanes of warm H$_{2}$ gas observed in NGC\,1068 and NGC\,1386 are wider than the narrow dust filaments and more extended, filling gaps in the dust morphology. This is also observed in other Seyfert galaxies like NGC\,7743 (\citealt{2014ApJ...792..101D}) and suggests that gas and dust trace \textit{two different phases} at scales close to the nuclear region: once dust and gas approach the central few 100 pc, the gas is first warmed up by the nucleus (and/or in some cases jet shocks; see Section~\ref{sourceh2}) and thus presents a wider and more extended morphology, while the dust remains a colder phase with a temperature well below 2000 K. This suggests that the dust morphology should resemble that of the cold molecular gas (e.g. CO at 10-40 K). The latter can be tested in those galaxies for which high angular resolution millimetre observations have been performed using ALMA (Cen\,A, \citealt{2013ASPC..476...69E}; NGC\,1068, \citealt{2014A&A...567A.125G}; NGC\,1097, \citealt{2013PASJ...65..100I}; NGC\,1566, \citealt{2014A&A...565A..97C}). In NGC\,1566, a dual spiral structure of cold CO(3-2) gas emission is observed to emerge up to 300 pc from the nuclear region and to indeed coincide with the dust extinction (Fig.~\ref{n1566}, bottom-left panel; see also \citealt{2014A&A...565A..97C}). Such a spiral structure is however not observed in our molecular H$_{2}$ warm gas emission map despite the large FoV of NaCo (Fig.~\ref{n1566}, bottom right panel). A good correlation between the cold CO(3-2) gas emission and dust is also observed for the Seyfert 2 galaxy NGC\,1433 (\citealt{2013A&A...558A.124C}) and for NGC\,1068 (\citealt{2014A&A...567A.125G}). 
The suggested \textit{two phases of accretion} in which dust remains colder than the gas when approaching the nuclear (few 100 pc) scales is also clearly illustrated in CenA: a spatial anticorrelation is observed between the warm H$_{2}$ and cold CO(2-1) gas, with the H$_{2}$ filling the central $\sim$2 arcsec region devoid of cold gas (Fig.~\ref{cena}, bottom right panel), while the cold CO(2-1) gas seems to extend in the same direction as the dust filament that comes from a few tens of pc to the nucleus (Fig.~\ref{cena}, bottom-left panel). In the case of CenA the warm H$_{2}$ gas distribution presents a shell-like structure that extends along the direction of the radio jet (see Fig.~\ref{cena}, middle-left panel), where no cold CO(2-1) gas emission is observed (Fig.~\ref{cena}, bottom panels). This warm H$_{2}$ gas morphology and the lack of cold gas emission in that direction are thus most likely caused by jet bow shocks (see Section~\ref{sourceh2}).

While the cold molecular gas distribution resembles the morphology of the dust but not that of the warm molecular gas, as predicted by the \textit{two phases of accretion} scenario and observed in all the target sources for which ALMA observations are available, the warm (1000-3000 K) and cold (10-40 K) molecular gas are part of the same dynamical structure and should thus show similar morphologies (or not be completely spatially decoupled) at 'mixed' or 'intermediate' temperatures of e.g. 40-500 K. A good illustration of this is observed in NGC\,1068 and NGC\,1097: the morphology of the warm H$_{2}$ gas closely resembles that of the slightly warm molecular gas ($T>$40 K) such as CO(6-5) or HCN, among others, as shown in the ALMA images of NGC\,1068 (Fig.~\ref{n1068}, bottom right panel) and NGC\,1097 (Fig.~\ref{n1097}, bottom-left panel; see also \citealt{2013PASJ...65..100I}, \citealt{2015A&A...573A.116M}), while, as expected from the \textit{two phases of accretion} scenario, the warm H$_{2}$ gas appears disconnected from the cold dust (e.g. Fig.~\ref{n1097}, middle-left panel). 


\subsection{Source of the warm H$_{2}$ molecular gas emission}
\label{sourceh2}
To explain the excitation of the warm H$_{2}$ gas, we look for evidence of spatially associated star formation, shocks produced by jet interaction or thermal heating by X-rays. Clumps of warm molecular H$_{2}$ gas emission coincident with regions of star formation in the \textit{HST} \textit{I}-band, NaCo/\textit{K}-band or ESO/VISIR mid-IR bands are observed in the northern spiral arm of Circinus (Fig.~\ref{circinus}, see Appendix\,\ref{individual}), the starburst ring of NGC\,1097 (Fig.~\ref{n1097}) and NGC\,7582 (Fig.~\ref{n7582}). In these regions, star formation is the heating mechanism that excites the H$_{2}$ molecules. Apart from these clumps, no warm H$_{2}$ gas emission is observed beyond scales of a few 100 pc despite the large FoV ($\sim$30 arcsec) of the H$_{2}$ intensity maps. We thus conclude that the H$_{2}$ maps obtained with NaCo show the majority of the warm H$_{2}$ molecular gas present in the studied galaxies.

No clumps of star formation are observed in the nuclear region of any of the sources (except for the northern spiral arm of Circinus mentioned above), indicating that the central warm H$_{2}$ gas is heated by the AGN. In nuclear regions, the H$_{2}$ molecules are found to be typically excited by thermal processes (e.g. \citealt{2004A&A...425..457R}; \citealt{2013MNRAS.428.2389M}; \citealt{2014A&A...565A..97C}) due to either shocks (\citealt{1989ApJ...342..306H}) or X-ray illumination (\citealt{1996ApJ...466..561M}). Most of the galaxies studied here were analysed in \cite{2003MNRAS.343..192R} using long-slit near-IR data. On the basis of the H$_{2}$ 2-1S(1)/1-0S(1) ratio, thermal excitation was favoured as the dominant mechanism for the nuclear H$_{2}$ observed in their sample of 14 Seyfert galaxies. In the present study, where we trace the whole distribution of warm H$_{2}$ within the central kpc region, we can state that no young star formation is seen associated with the central but spatially extended H$_{2}$ emission in any of the galaxies: no evidence for it is seen at the centre in our high angular resolution \textit{K}-band images. UV photons are thus not the cause of the source of H$_{2}$ excitation, in agreement with the \cite{2003MNRAS.343..192R} conclusion.

We find, however, a fair spatial correlation between the H$_{2}$ and the X-ray morphology, which is particularly evident in those cases where the X-ray and H$_{2}$ emission is seen spatially resolved at the centre: for NGC\,1068 and NGC\,1386, the nuclear X-ray emission presents an extension and morphology that resembles that of the warm H$_{2}$ molecular gas (middle-right panel in Fig.~\ref{n1068}, for NGC\,1068, and bottom-right panel Fig.~\ref{n1386}, for NGC\,1386) and is coincident with the extent and morphology of the narrow-line region traced by the [OIII] ionized gas (\citealt{1997ApJ...476L..67C}; \citealt{2006A&A...448..499B}; \citealt{2012ApJ...756..180W}), suggesting that both the diffuse X-ray and the [O III] emission are produced by the same gas photoionized by the nuclear continuum (e.g. \citealt{2003MNRAS.345..369I}; \citealt{2012ApJ...756..180W}). The extension of the nuclear 0.3-10 keV band X-ray emission is also comparable to the nuclear warm H$_{2}$ gas distribution in NGC\,1097, NGC\,7582 and Circinus (Figs.~\ref{circinus}, \ref{n1097}, \ref{n7582}, bottom right panel) indicating that the nuclear warm H$_{2}$ molecular gas is most likely excited by the X-ray emission of the photoionized gas, as it was also suggested for e.g. the Seyfert galaxies NGC\,4151, NGC\,5548 and NGC\,3227 (\citealt{2004A&A...425..457R}; \citealt{2009MNRAS.394.1148S}). The nuclear X-ray emission traced by the \textit{Chandra} X-ray satellite for CenA (\citealt{2004ApJ...617..209E}) is observed to cover an area larger than the extent of the nuclear warm H$_{2}$ molecular gas. For NGC\,1566, the X-ray emission is unresolved (\citealt{1992ApJS...80..531F}) and does not allow a direct morphological comparison with the warm H$_{2}$ gas distribution. 

Hard X-ray photons can easily penetrate the molecular clouds before being absorbed and heat the medium to excite the 2$\mu$m vibration-rotation lines of H$_{2}$. We test this possibility from the energy balance point of view by comparing the observed H$_{2}$ fluxes with those predicted in the models of \cite{1996ApJ...466..561M}. These models investigate the effects of X-ray irradiation on molecular gas and predict H$_{2}$ fluxes for a wide range of X-ray impinging luminosities and densities. We follow a similar approach as in \cite{2004A&A...425..457R}. Accordingly, we determine the effective ionization parameter, defined in \cite{1996ApJ...466..561M} as follows:
\begin{equation}
\zeta \simeq 100 \frac{L_\mathrm{X}}{n_\mathrm{e}d^{2}N_\mathrm{H_{att}}^{0.9}}
\end{equation}
where $L_\mathrm{X}$ is the hard X-ray luminosity in units of 10$^{44}$ erg s$^{-1}$, $n_\mathrm{e}$ the total gas density in units of 10$^{5}$ cm$^{-3}$, $d$ the distance in pc of the emitting gas to the AGN, and $N_\mathrm{H_{att}}$ the attenuating column density to the X-rays in units of $10^{22}$ cm$^{-2}$, and we use Fig. 6 in \cite{1996ApJ...466..561M} assuming $n_\mathrm{e}$=$10^{5}$ cm$^{-3}$.
The X-ray nuclear luminosities are taken for most sources from \cite{2010MNRAS.402..724P}, who compile fluxes at energies above 20 keV. These reflect the truly X-ray emerging flux from the nucleus, particularly for the Compton thick sources in the sample. For NGC\,1386, we use the 2--10 keV X-ray flux from \cite{1997PASJ...49..425I}. The average X-ray luminosity of the sample is $\sim10^{42}$ erg s$^{-1}$ (see Table~\ref{table2}). The attenuating $N_\mathrm{H}$ correspond to the $A_\mathrm{v}$ extinction values derived by \cite{2002MNRAS.331..154R} and \cite{2005AJ....130.1472P,2010MNRAS.402..724P,2014MNRAS.442.2145P} for the central few 100 pc in these galaxies, which is within the radius the H$_{2}$ emission is observed. We used the conversion factor $N_\mathrm{H}$= (1.79 $\pm$ 0.03) $\times$ 10$^{21}$ cm$^{-2}$ $A_\mathrm{v}$ (\citealt{1995A&A...293..889P}). The values of $N_\mathrm{H_{att}}$ obtained are typically in the range 10$^{21}$--10$^{22}$ cm$^{-2}$ (see Table~\ref{table4}). The radius of the nuclear extended H$_{2}$ emission is taken from Table~\ref{table2}. 

\begin{table}
\centering
\scriptsize
\caption{Predicted H$_{2}$ 2.12$\mu$m fluxes from the X-ray models of \citet{1996ApJ...466..561M} for a gas density of 10$^{5}$ cm$^{-3}$.}
\label{table4}
\begin{tabular}{lcccc}
\hline
\hline 
			&					&					\multicolumn{3}{c}{H$_2$ 1--0 S(1) flux} 								 \\
Object	     	&   $N_\mathrm{H_{att}}$  	&     Predicted   			& Observed  			&	Predicted if f=10\%			 \\
	       		&      [10$^{22}$ cm$^{-2}$]   &   [erg cm$^{-2}$ s$^{-1}$] 	&[erg cm$^{-2}$ s$^{-1}$]		&	[erg cm$^{-2}$ s$^{-1}$]			\\
\hline
CenA         	&  2.4	 			&	$1.2\times10^{-14}$	&	$1.6\times10^{-14}$			&	$1.2\times10^{-13}$				\\ 
Circinus		& 1.3 	 			&	$3.9\times10^{-15}$	&	$1.9\times10^{-14}$			&	$3.9\times10^{-14}$				\\
NGC\,1068	&  0.4	 			&	$4.5\times10^{-15}$	&	$7.1\times10^{-14}$			&	$4.5\times10^{-14}$				\\
NGC\,1097	&  0.04	 			&	$6.0\times10^{-15}$	&	$5.4\times10^{-14}$			&	$6.0\times10^{-14}$				\\			
NGC\,1386	&  0.5	 			&	$1.4\times10^{-16}$	&	$6.6\times10^{-14}$			&	$1.4\times10^{-15}$				\\
NGC\,1566   	&  0.6	 			&	$9.1\times10^{-15}$	&	$8.9\times10^{-14}$			&	$9.1\times10^{-14}$				\\
NGC\,7582	&  0.9	 			&	$2.1\times10^{-16}$	&	$3.4\times10^{-15}$			&	$2.1\times10^{-15}$				\\  
\hline
\end{tabular}
\raggedright
\smallskip\newline\small 
$^*$ Predicted H$_{2}$ 2.12$\mu$m fluxes assuming a conservative filling factor of 10\%.
\end{table}

The predicted H$_{2}$ flux apparently agrees well with the one measured only for CenA. However, considering the uncertainties in the determination of the ionization parameter, which could be up to an order of magnitude higher if allowed for lower $N_\mathrm{H_{att}}$, the predicted H$_{2}$ fluxes are within the range of observed values also for Circinus, NGC\,1097 and NGC\,1566 (see Table~\ref{table4}). In the remaining three, the prediction is an order of magnitude less, even if $N_\mathrm{H_{att}}$ is allowed to get reduced by an order of magnitude --this is because the dependence of the H$_{2}$ flux on $\zeta$ is weak for the range of $\zeta$ derived for these sources. Also much lower $N_\mathrm{H}$ would expose easily the core of the H$_{2}$ molecular cloud to the X-rays and would become fully ionized.  
The underpredicted H$_{2}$ fluxes are thus surprising given the spatial association between the H$_{2}$ and the X-rays. Yet, we know that the interstellar medium in galaxies is very clumpy, with molecular gas clouds having average sizes in the 10-100 pc range, and filling factors of a few percent are estimated for the cold dense phase (\citealt{1977ApJ...218..148M}). A way to reconcile the above results is thus to assume a filling factor in the molecular gas region so that the observed emission arises from the most clumpy regions inside the molecular gas region that have higher densities; this would also explain the relatively large radii at which H$_{2}$ emission is detected. Our knowledge of the filling factor of molecular clouds in the environment of AGN is poor, yet assuming filling factors ($f$) in the range of 1\% to 10\%, in line with the values adopted by \cite{1977ApJ...218..148M} in their three-component medium of the interstellar medium, would bring the results to agreement (Table~\ref{table4}, last column). An exception is NGC\,1386, where the H$_{2}$ emission is the largest among the objects in the sample. This source has a jet (\citealt{1999ApJ...516...97N}) of size$\sim$100 pc and power 1.2$\times10^{42}$ erg s$^{-1}$ (\citealt{2014ApJ...787...62M}) pointing to the south direction and coinciding in location with the southern extension in H$_{2}$, thus shocks induced by the jet might in this case be the local source of H$_{2}$ excitation.

Shock excitation of the H$_{2}$ molecule caused by jet interactions could also take place in CenA and NGC\,1097, who have a radio jet of total size $\sim$2 kpc and $\sim$0.1 kpc, respectively, and values of jet power ($6\times10^{42}$ erg s$^{-1}$ for CenA and $1.5\times10^{41}$ erg s$^{-1}$ for NGC\,1097; e.g. \citealt{2014ApJ...787...62M} and references therein) several orders of magnitude larger than the H$_{2}$ luminosity. For CenA, this is supported by the well-defined shell-like structure of the warm H$_{2}$ molecular gas emission, which extends perpendicular to the direction of the radio jet, and by the finding of a large H$_{2}$/PAH (polycyclic aromatic hydrocarbon) ratio, which rules out UV excitation in photodissociation regions as the source of the H$_{2}$ line excitation (\citealt{2010ApJ...724.1193O}). Therefore, although CenA is the only source in the sample that is comfortably explained by thermal excitation by X-rays (see Table~\ref{table4}, column 4), the bow shock morphology of the H$_{2}$ gas emission and apparent fair alignment between the H$_{2}$ gas and the jet direction qualitatively points to shocks induced by the jets as a plausible additional mechanism for exciting the H$_{2}$ gas. In NGC\,1068, a recent episode of star formation ($\sim$30 Myr ago; \citealt{2012ApJ...755...87S}) was found to occur in the ring-like structure of warm H$_{2}$ gas located at $\sim$100 pc from the nucleus. According to simulations, the formation of such rings and star formation at radii of $\sim$ 100 pc could result from the shock waves created by a young radio jet (\citealt{2012MNRAS.425..438G}). The excitation of warm H$_{2}$ molecular gas ring-like structure in NGC\,1068 could thus be also explained by a shock jet interaction (see Fig.~\ref{n1068}, middle-left panel). In this case, the off-centred nuclear position of the ring could be explained by inhomogeneities in the gas that surrounds the nucleus (\citealt{2014MNRAS.445.2353B}).


The large extent of the nuclear warm H$_{2}$ gas, up to scales of a few 100 pc, can be thus explained for all the targets by thermal processes, in agreement with previous studies (e.g. \citealt{2004A&A...425..457R}; \citealt{2013MNRAS.428.2389M}; \citealt{2014A&A...565A..97C}). Given the similar symmetrical morphology between the nuclear X-ray and warm H$_{2}$ molecular gas emission, X-ray heating is the most likely excitation source of the nuclear warm H$_{2}$ gas for most sources, possibly also combined with shocks from radio jets for CenA, NGC\,1097 and NGC\,1068. 

Finally, we note that the nuclear X-ray luminosity of the target sources is one to four orders of magnitude larger than the nuclear H$_{2}$ luminosity (see Table~\ref{table2}), indicating that a small fraction of X-ray photons penetrating and heating the central H$_{2}$ molecular gas cloud is enough to excite the H$_{2}$ gas. The X-ray luminosities of the AGN studied here are also two to four orders of magnitude higher than those of the quiescent galaxies of \cite{2013ApJ...768..107H} (see also \citealt{2013MNRAS.428.2389M}), which show not so centrally concentrated nuclear warm H$_{2}$ gas distributions and lower H$_{2}$ luminosities than the AGN, thus reinforcing the AGN X-ray emission as the dominant excitation mechanism of the warm H$_{2}$ molecular gas.

\subsection{Molecular gas mass and density}
\label{mass}
The large FoV of the NaCo observations reveals the extent and morphology of all the warm molecular H$_{2}$ gas present in the studied galaxies at scales of a few 100 pc. Following \cite{2002MNRAS.331..154R}, the mass of this warm H$_{2}$ gas ($M_\mathrm{warm}$) is quantified using the flux extracted from the H$_{2}$ intensity maps and assuming a gas temperature of 2000 K, the H$_{2}$ 1--0S(1) transition probability $A_{S(1)}$ = 3.47 $\times$ 10$^{-7}$ s$^{-1}$ (\citealt{1977ApJS...35..281T}) and the population fraction of the $(\nu,J)$ = (1,3) level $f_\mathrm{(1,3)}$ = 0.0122 (\citealt{1982ApJ...253..136S}) as:
\begin{equation}
\small
M_\mathrm{warm} = 5.0875 \times 10^{13} (\frac{D_{L}}{\mathrm{Mpc}})^{2} \frac{F_{1-0S(1)}}{\mathrm{erg\ s^{1} cm^{-2}}} 10^{0.4277 A_{2.2}}
\end{equation}
where $A_{2.2}$ is the extinction at 2.2 $\mu$m (\citealt{1982ApJ...253..136S}). The H$_{2}$ flux was extracted in a nuclear circular aperture for CenA, Circinus, NGC\,1068, NGC\,1097, NGC\,1566 and NGC\,7582, and an elliptical aperture for NGC\,1068 and NGC\,1386 with a size corresponding to the extent of the warm H$_{2}$ gas emission given by the isophote fitting (Table~\ref{table2}). The values of $A_{2.2}$, derived from regions surrounding the nucleus, are taken from \cite{2002MNRAS.331..154R} for CenA and NGC\,1566, \cite{2005AJ....130.1472P} for NGC\,1097, \cite{2010MNRAS.402..724P} for Circinus and NGC\,1068 and \cite{2014MNRAS.442.2145P} for NGC\,1386 and NGC\,7582, and range between $A_{2.2}\ =$ 0.2 and 1.6 for all sources. The masses of the nuclear warm H$_{2}$ gas are reported in Table~\ref{table2} and range between 40 M$_{\odot}$ (for Circinus) and 3000 M$_{\odot}$ (for NGC\,1566). 

In Section~\ref{morphology} we found that the nuclear H$_2$ luminosity of the sample is on average 10 times higher than that of the quiescent sources from \cite{2013ApJ...768..107H} and \cite{2013MNRAS.428.2389M}. The mass of the AGN molecular gas reservoir is thus $\sim$ 10 times higher for the Seyfert galaxies studied here than for the quiescent galaxies (as found by \citealt{2013ApJ...768..107H}). This suggests that either these AGN have more material available for accretion on to the central SMBH than the non-AGN galaxies, which could also explain the higher X-ray luminosities of the Seyfert galaxies, or that the gas is more effectively heated by the AGN in the Seyfert galaxies, so that in these we see the H$_{2}$ emission while in the normal galaxies the H$_{2}$ gas remains invisible. 

The total content of molecular cold gas can be estimated using a correlation between H$_{2}$ 2.12 $\mu$m luminosity and cold gas mass derived from CO observations (e.g. \citealt{2005AJ....129.2197D}; \citealt{2006A&A...454..481M}; \citealt{2013MNRAS.428.2389M}). This allows us to obtain an estimate of the amount of cold gas present in the nuclear regions and thus available for accretion on to the AGN. Using the $M_\mathrm{cold}$--$L_\mathrm{H2}$ relation of \cite{2013MNRAS.428.2389M}, equation\,4, we obtain that the cold H$_{2}$ gas mass ranges between $7\times10^{6}\, \rm{M_{\odot}}$ (for CenA) and $1\times10^{9}\, \rm{M_{\odot}}$ (for NGC\,1566; see Table~\ref{table2}). The value of $M_\mathrm{cold}$ obtained for NGC\,1566 is in agreement with the total molecular mass of $1.3\times10^{9}\, \rm{M_{\odot}}$ obtained from CO Swedish-ESO Submillimetre Telescope (SEST) observations but one order of magnitude lower than the $M_\mathrm{cold}$ obtained from smaller FoV CO observations (e.g. 18 arcsec with ALMA; see \citealt
 {2014A&A...565A..97C} for a discussion), which indicates that NaCo recovers the total content of molecular gas for NGC\,1566. For CenA, Circinus and NGC\,1068, the estimated values of $M_\mathrm{cold}$ are one order of magnitude lower than those obtained from CO observations (\citealt{1998A&A...338..863C}; \citealt{2000ApJ...533..850S}; \citealt{2009ApJ...695..116E}), which can be mainly attributed to the larger FoV of the CO observations. We note though that better agreement would not be achieved even if we considered a larger aperture for the warm H$_{2}$ gas, as (except for the clumps in the spiral arm of NGC\,1097) most of the detected warm H$_{2}$ gas is concentrated in the nuclear regions and no emission is observed further out despite the large FoV of NaCo.

Given that the morphology of the nuclear warm H$_{2}$ molecular gas appears diffuse and symmetrical in most sources, we can estimate its density assuming a spherical volume with a gas mass fraction (or gas-to-mass ratio) $f_\mathrm{g}$=1\% for NGC\,1097 and $f_\mathrm{g}$=10\% for the rest of the sources (following \citealt{2009ApJ...696..448H}). This $f_\mathrm{g}$ represents the fraction of the volume occupied by a certain gas phase and should not be confused with the filling factor ($f$) introduced in Section~\ref{sourceh2}\footnote{\cite{2009ApJ...696..448H} assumed a uniform gas distribution in their work (this is, a filling factor of 100\%) though the authors conclude that the gas must be clumpy. This implies a filling factor $f<$1, in line with the hypothesis made in Section~\ref{sourceh2} of assuming a filling factor for the H$_{2}$ molecular gas.}. The volume is derived using the same radius as that considered when calculating the nuclear H$_{2}$ gas masses. We obtain cold gas densities ranging $N_\mathrm{cold}$= $6\times10^{3}--1\times10^{5}$ cm$^{-3}$, which is consistent with the $N_\mathrm{H_{2}}$= 10$^{4}$--10$^{5}$ cm$^{-3}$ typically derived from CO observations (e.g. \citealt{1989A&A...223...79M}; \citealt{2009ApJ...707..126B}).

\section{Conclusions}
\label{conclusions}
In this paper we have presented the morphology and properties of the warm molecular H$_2$ gas of a sample of seven nearby Seyfert galaxies observed with VLT AO and narrow-band imaging with NaCo. Dust colour maps of these galaxies are also constructed using the ratio between broad-band \textit{K}-band images and \textit{V}- and \textit{I}-band \textit{HST} images. The large FoV of the observations has allowed us to trace the distribution of the warm H$_2$ gas and dust up to scales of a few $100\, \rm{pc}$ with a resolution $<0.5$ arcsec. 

For all galaxies, the warm H$_2$ gas is concentrated in the nuclear regions (with an HWHM of 27 pc) and does not extend beyond $\sim 300\, \rm{pc}$ from the nucleus. The warm H$_{2}$ gas is thus visible when the gas is close enough to the heating source (the nucleus, where the peak of H$_{2}$ emission is observed) and the H$_{2}$ molecules are excited enough to produce ro-vibrational emission lines. Further out, the H$_{2}$ gas is not warm enough and only the cold gas is visible through e.g. CO observations. In NGC\,1097, Circinus and NGC\,7582 some clumps or lanes of H$_2$ emission are in addition observed to coincide with regions of star formation. 

The excitation mechanism for all the target sources is found to be thermal. The cospatial location of H$_2$ and X-rays, also coinciding on physical extension in those cases where spatial resolution allows for size measurements, is a strong indication for nuclear X-rays being the major source of excitation of H$_2$. We conclude that irradiation by X-rays is the predominant source of excitation for most sources under reasonable assumptions of the H$_2$ clouds volume filling factor (of 10\%). 

In NGC\,1068, NGC\,1386, NGC\,1566 and Circinus, the morphology of the warm H$_2$ gas traced either by NaCo or IFU observations resembles within a few tens of pc that of the filamentary dust lanes that extend from the central parsecs to $\sim 1\, \rm{kpc}$ scale. These dust filaments are potential fuelling channels that transport the gas to the nuclear region. This suggests that the nuclear warm H$_2$ gas of these galaxies comes from outer (kpc-scale) regions following these same channels of inflow and warms up as it approaches the nucleus. The observed lanes of warm H$_2$ gas are however much wider and extended than the filaments of dust, which suggests that gas and dust trace \textit{two different states of accretion with different temperatures}. In such scenario the nuclear dust morphology at scales of a few 100 pc should resemble that of the cold (10-40 K) molecular gas, as we observe for CenA and NGC\,1566 by ALMA CO observations, while the inner warm (1000-3000 K) H$_2$ gas presents a wider morphology that should resemble that of the slightly warmed gas (40-500 K), as indicated by warmer molecular tracers (e.g. CO(6-5), HCN) observed in NGC\,1068 and NGC\,1097 with ALMA.

\section{Acknowledgements}

The authors thank F. Israel for comments on the manuscript. The authors are indebted to F. M\"uller-S\'anchez for providing the SINFONI H$_2$ gas emission map of NGC\,1068, O. Gonz\'alez-Mart\'in for the \textit{Chandra} data, J. Wang for the \textit{Chandra} image of NGC\,1068, L. Burtscher for the SINFONI data cubes, F. Combes for the ALMA image of NGC\,1566, D. Espada for the ALMA image of CenA, and S. Garc\'ia Burillo for the ALMA image of NGC\,1068. MM acknowledges financial support from AYA2011-25527 and NASA \textit{Chandra} grant G05-16099X. The research leading to the results has received funding from the European Community's Seventh Framework Programme under grant agreement 312430. Research by JAFO was supported by Canary Islands CIE: Tricontinental Atlantic Campus. This research is based on ESO VLT programmes P71.B-0632A, 74.B-0722B and 86.B-0484. The research has made use of the NASA/IPAC Extragalactic Database (NED) which is operated by the Jet Propulsion Laboratory, California Institute of Technology, under contract with the National Aeronautics and Space Administration.

\bibliographystyle{mn2e} 
\bibliography{referencesALL}

\begin{figure*}
 \includegraphics[width=0.85\textwidth]{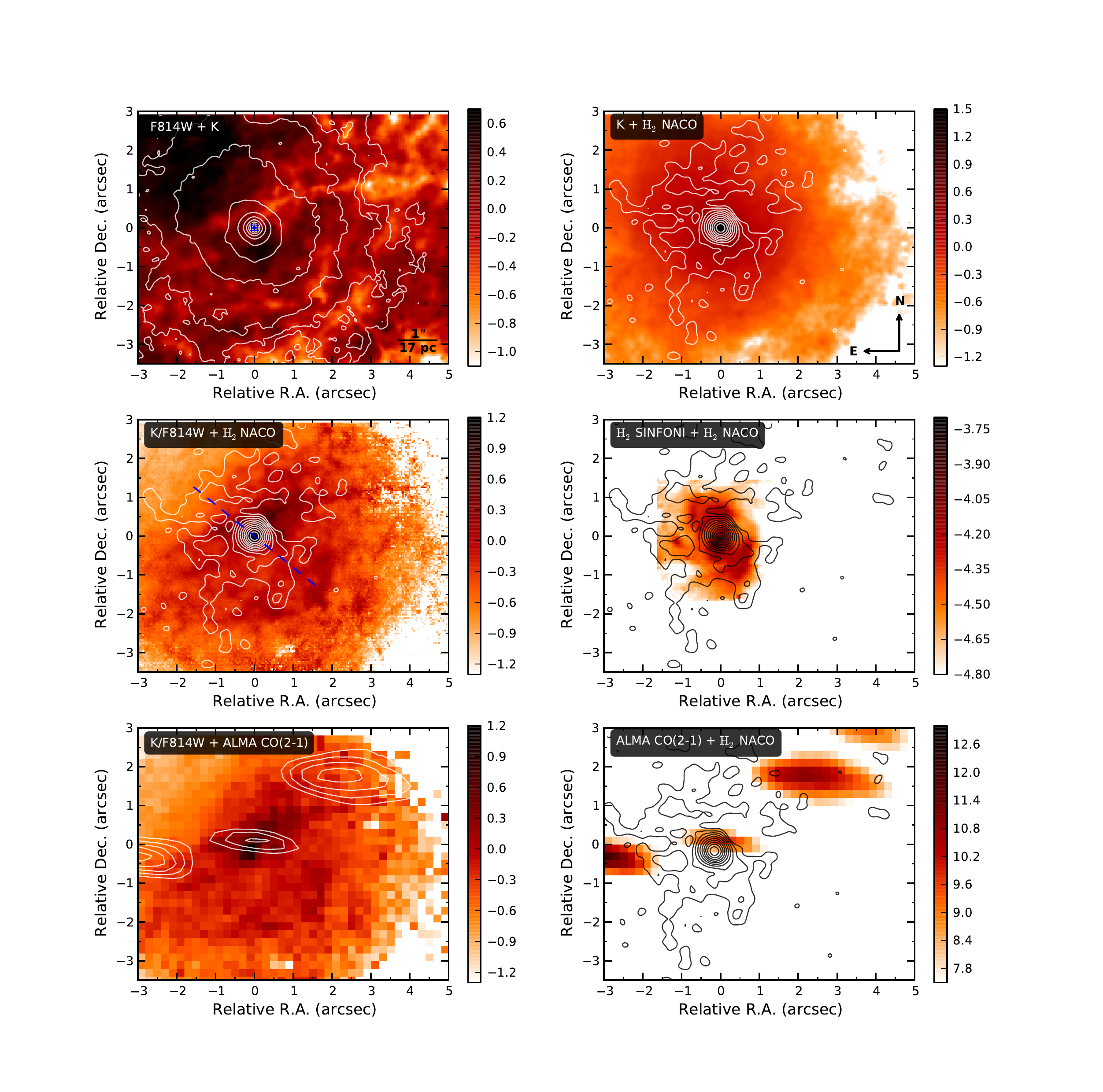}
 \protect\caption[cena]{CenA. \textbf{Top left}: \textit{HST/F814W} image with NaCo/\textit{K}-band continuum contours in white. The position of the nucleus and its error is marked with a cross. \textbf{Top right:} NaCo/\textit{K}-band image. The contours of the warm H$_{2}$ molecular gas obtained from NaCo are shown in white. They start at three times the background noise level and increase in factors of 2. \textbf{Middle left:} \textit{K/F814W} ratio or dust map. The contours of the warm H$_{2}$ molecular gas obtained from NaCo are shown in white. The orientation of the radio jet (e.g. \citealt{1992ApJ...395..444C}) is indicated with a dashed blue line. \textbf{Middle right:} Intensity map of the warm H$_{2}$ molecular gas obtained from IFU SINFONI. The contours of the warm H$_{2}$ molecular gas obtained from NaCo are shown in black. \textbf{Bottom left:} \textit{K/F814W} ratio or dust map. The contours of the cold CO(2-1) molecular gas obtained from ALMA (\citealt{2013ASPC..476...69E}) are shown in white. Contours start at three times the rms noise of 2.5 mJy beam$^{-1}$ (\citealt{2013ASPC..476...69E}). \textbf{Bottom right:} ALMA CO(2-1) integrated intensity map (from \citealt{2013ASPC..476...69E}). The contours of the warm H$_{2}$ molecular gas obtained from NaCo are shown in black. The colour scale is linear and starts at three times the rms noise of 2.5 mJy beam$^{-1}$. The colour palettes in the other panels are in arbitrary units. The FoV for all panels is 8 arcsec $\times$ 6 arcsec. North is up and east is to the left.}
 \label{cena}
\end{figure*}

\begin{figure*}
 \includegraphics[width=0.85\textwidth]{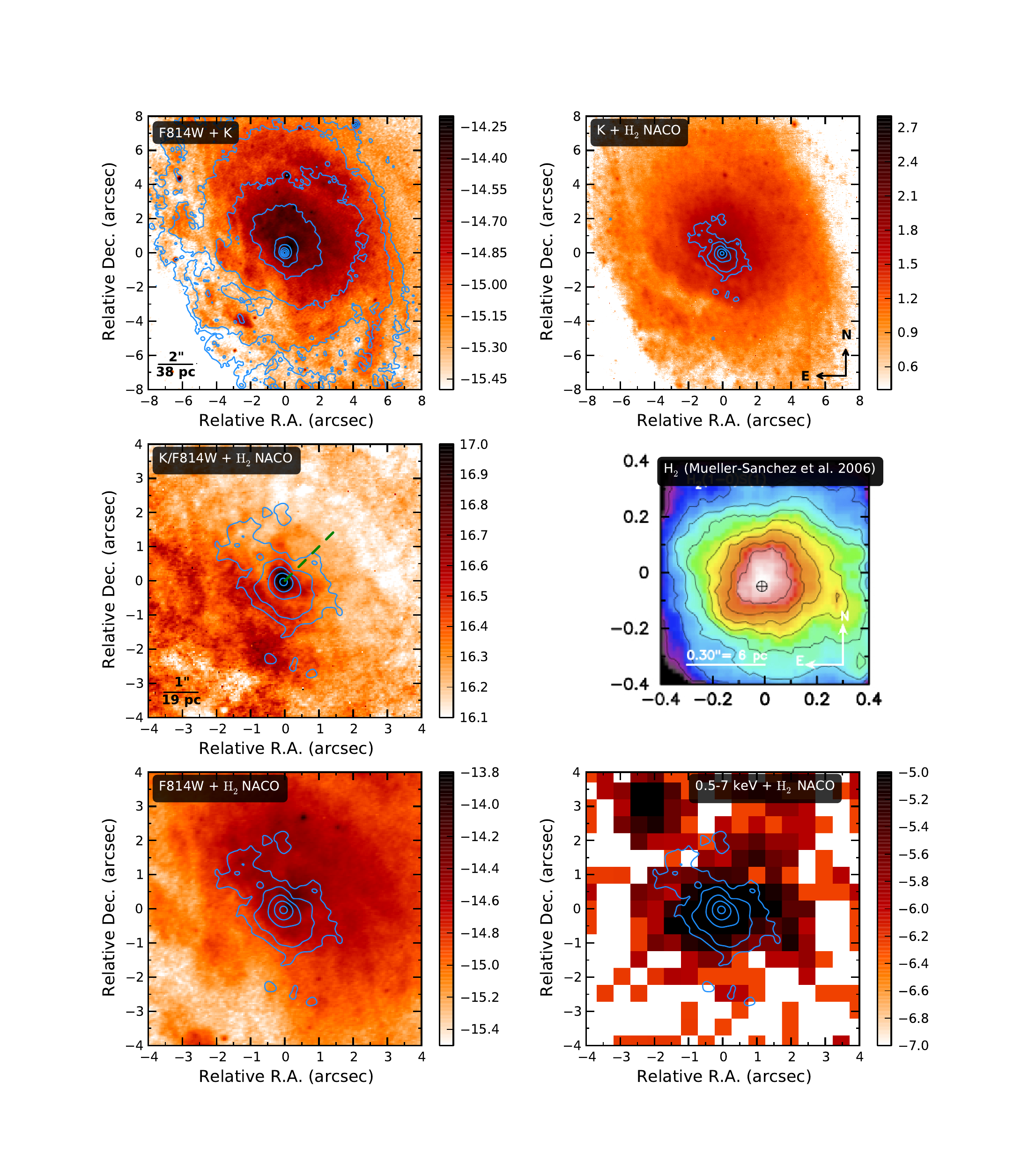}
 \protect\caption[circinus]{Circinus. \textbf{Top left}: \textit{HST/F814W} image with NaCo/\textit{K}-band continuum contours in blue. The FoV is 16 arcsec $\times$ 16 arcsec. \textbf{Top right:}  NaCo/\textit{K}-band image with the same FoV as the previous panel. The contours of the warm H$_{2}$ molecular gas obtained from NaCo are shown in blue. They start at three times the background noise level and increase in factors of 3. \textbf{Middle left:} \textit{K/F814W} ratio or dust map with a FoV of 8 arcsec $\times$ 8 arcsec. The contours of the warm H$_{2}$ molecular gas obtained from NaCo are shown in blue. The orientation of the narrow line region (e.g. \citealt{1994A&A...291...18M}) is indicated with a dashed green line. \textbf{Middle right:} Intensity map of the H$_{2}$ molecular gas obtained from IFU SINFONI (from \citealt{2006A&A...454..481M}). \textbf{Bottom left:} \textit{HST/F814W} image with a FoV of 8 arcsec $\times$ 8 arcsec. The contours of the warm H$_{2}$ molecular gas obtained from NaCo are shown in blue. \textbf{Bottom right:} \textit{Chandra} X-ray image in the 0.5-7 keV band with the same FoV as the previous panel. The contours of the warm H$_{2}$ molecular gas obtained from NaCo are shown in blue. colour scales are linear and in arbitrary units. North is up and east is to the left.}
 \label{circinus}
\end{figure*}

\begin{figure*}
 \includegraphics[width=0.85\textwidth]{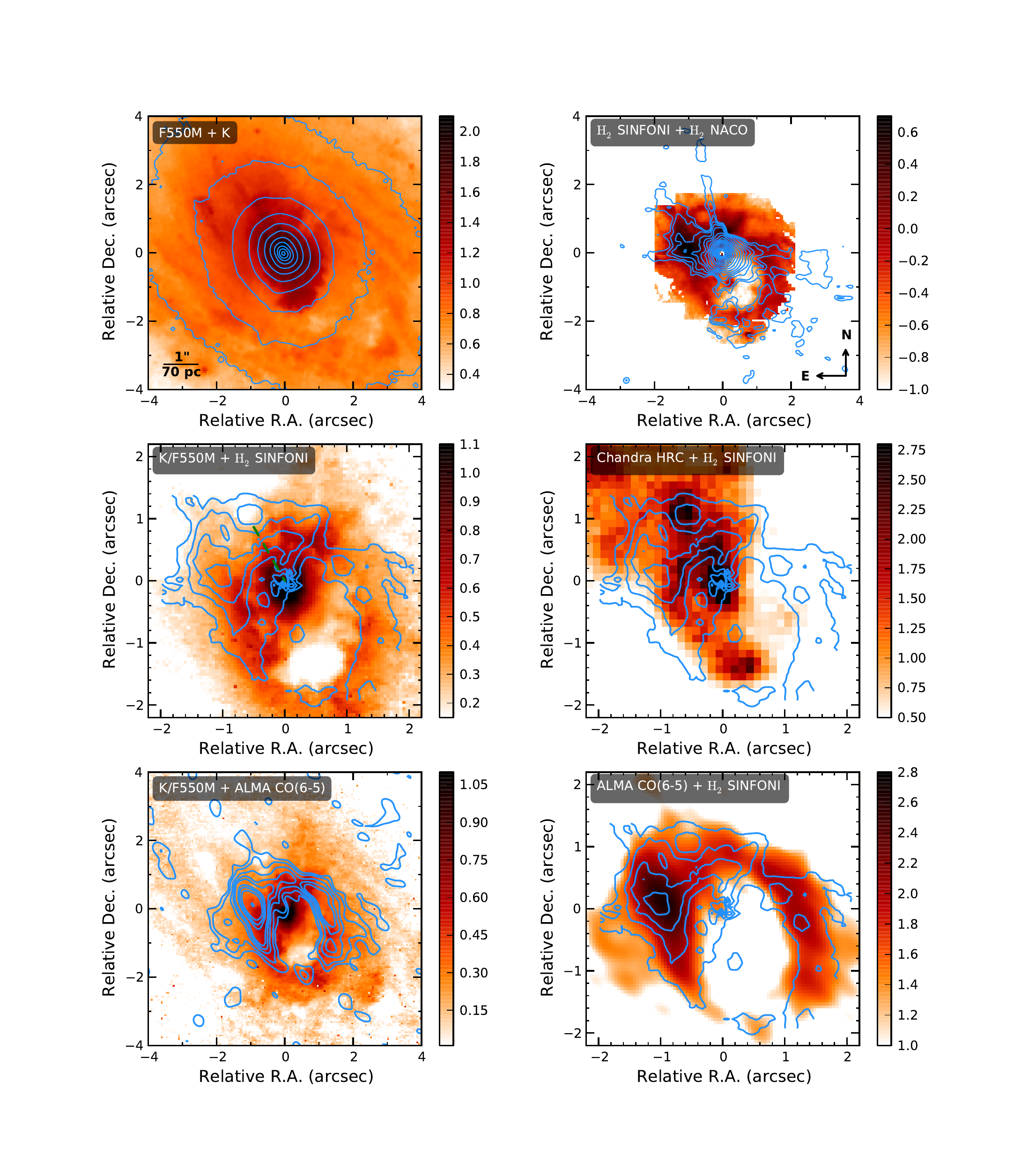}
 \protect\caption[n1068]{NGC\,1068. \textbf{Top left}: \textit{HST/F550M} image with NaCo/\textit{K}-band continuum contours in blue. The FoV is 8 arcsec $\times$ 8 arcsec. \textbf{Top right:} Intensity map of the warm H$_{2}$ molecular gas obtained from IFU SINFONI (from \citealt{2009ApJ...691..749M}). The FoV is the same as the previous panel. The contours of the warm H$_{2}$ molecular gas obtained from NaCo are shown in blue. They start at three times the background noise level and increase in factors of 3. \textbf{Middle left:} \textit{K/F550M} ratio or dust map with a FoV of 4.4 arcsec $\times$ 4.4 arcsec. The contours of the warm H$_{2}$ molecular gas obtained from IFU SINFONI are shown in blue. The orientation of the radio jet (e.g. \citealt{1996ApJ...458..136G}) is indicated with a dashed green line. \textbf{Middle right:} \textit{Chandra} HRC image from \cite{2012ApJ...756..180W} with the same FoV as the previous panel. The contours of the warm H$_{2}$ molecular gas obtained from SINFONI are shown in blue. \textbf{Bottom left:} \textit{K/F550M} ratio or dust map with a FoV of 8 arcsec $\times$ 8 arcsec. The contours of the CO(6-5) molecular gas obtained from ALMA (\citealt{2014A&A...567A.125G}) are shown in blue. Contours levels go from 5$\sigma$ to 240$\sigma$, where $\sigma$ =  2 Jy beam$^{-1}$ km s$^{-1}$ (\citealt{2014A&A...567A.125G}). \textbf{Bottom right:} ALMA CO(6-5) integrated intensity map (from \citealt{2014A&A...567A.125G}) with a FoV of 4.4 arcsec $\times$ 4.4 arcsec. The contours of the warm H$_{2}$ molecular gas obtained from SINFONI are shown in blue. The colour scale is logarithmic in 0.87 Jy beam$^{-1}$ km s$^{-1}$. The colour palettes in the other panels are logarithmic and in arbitrary units. North is up and east is to the left.}
 \label{n1068}
\end{figure*}

\begin{figure*}
 \includegraphics[width=0.85\textwidth]{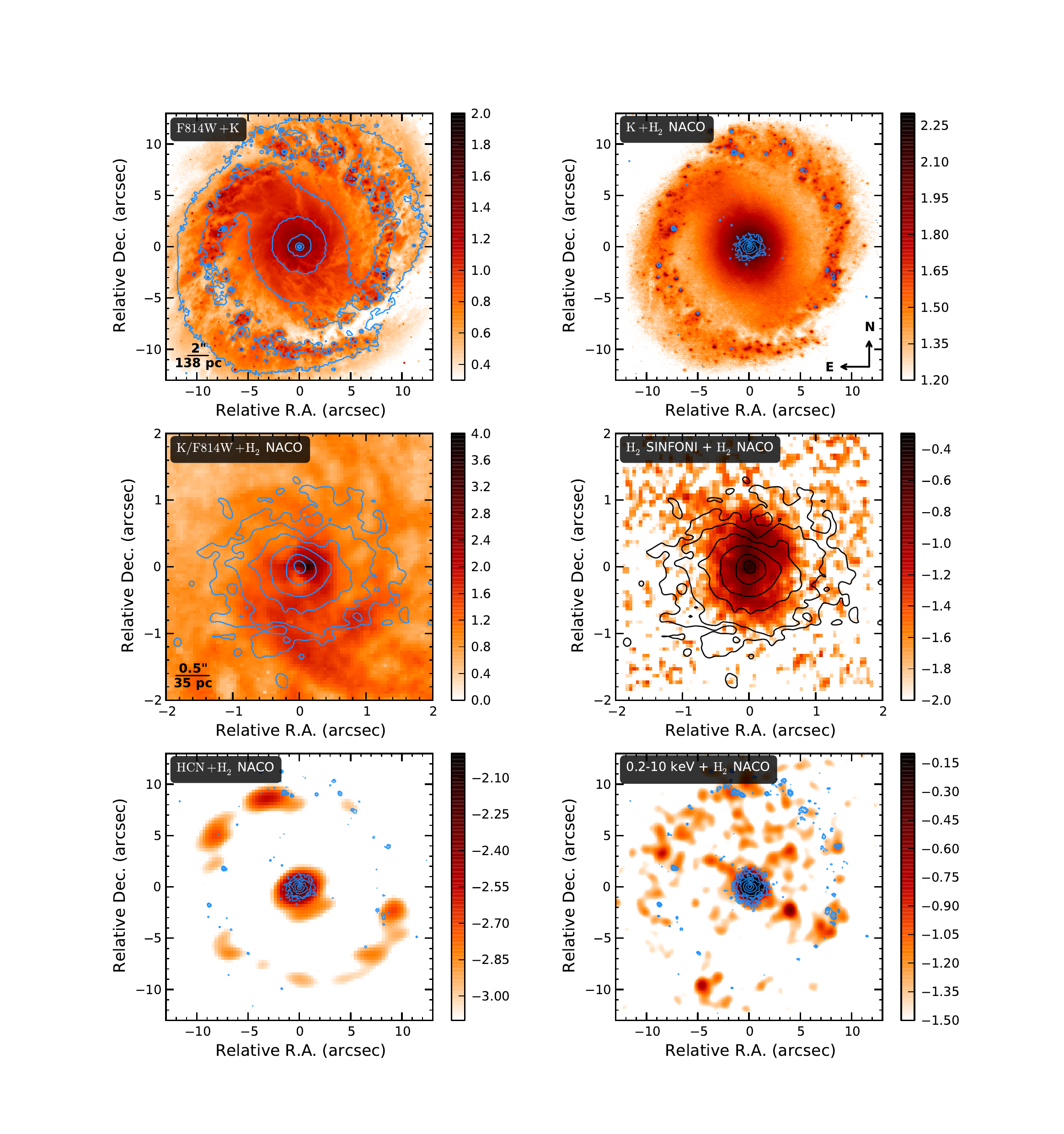}
 \protect\caption[n1097]{NGC\,1097. \textbf{Top left}: \textit{HST/F814W} image with NaCo/\textit{K}-band continuum contours in blue. The FoV is 26 arcsec $\times$ 26 arcsec. \textbf{Top right:} NaCo/\textit{K}-band image with the same FoV as the previous panel. The contours of the warm H$_{2}$ molecular gas obtained from NaCo are shown in blue. They start at three times the background noise level and increase in factors of 2. \textbf{Middle left:} \textit{K/F814W} ratio or dust map with a FoV of 4 arcsec $\times$ 4 arcsec. The contours of the warm H$_{2}$ molecular gas obtained from NaCo are shown in blue. \textbf{Middle right:} Intensity map of the warm H$_{2}$ molecular gas obtained from IFU SINFONI. The FoV is the same as the previous panel. The contours of the warm H$_{2}$ molecular gas obtained from NaCo are shown in black. \textbf{Bottom left:} HCN image with a FoV of 26 arcsec $\times$ 26 arcsec. The contours of the warm H$_{2}$ molecular gas obtained from NaCo are shown in blue. \textbf{Bottom right:} \textit{Chandra} X-ray image in the 0.2-10 keV band with the same FoV as the previous panel. The contours of the warm H$_{2}$ molecular gas obtained from NaCo are shown in blue. colour scales are logarithmic and in arbitrary units. North is up and east is to the left.}
 \label{n1097}
\end{figure*}

\begin{figure*}
 \includegraphics[width=0.85\textwidth]{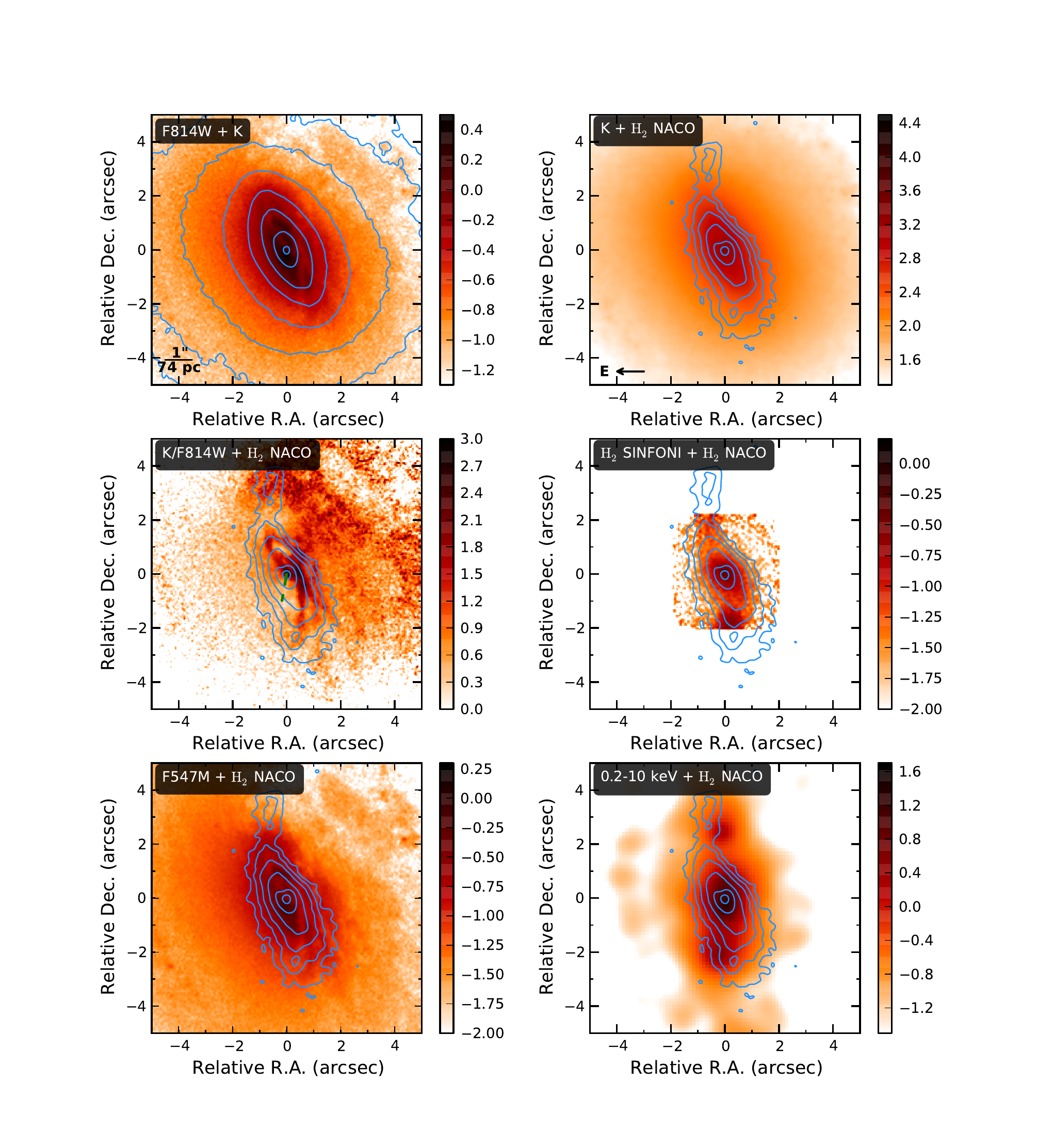}
 \protect\caption[n1386]{NGC\,1386. \textbf{Top left}: \textit{HST/F814W} image with NaCo/\textit{K}-band continuum contours in blue.  \textbf{Top right:}NaCo/\textit{K}-band image. The contours of the warm H$_{2}$ molecular gas obtained from NaCo are shown in blue. They start at three times the background noise level and increase in factors of 2. \textbf{Middle left:} \textit{K/F814W} ratio or dust map. The contours of the warm H$_{2}$ molecular gas obtained from NaCo are shown in blue. The orientation of the radio jet (e.g. \citealt{1999ApJ...516...97N}) is indicated with a dashed green line. \textbf{Middle right:} Intensity map of the warm H$_{2}$ molecular gas obtained from IFU SINFONI. The contours of the warm H$_{2}$ molecular gas obtained from NaCo are shown in blue. \textbf{Bottom left:} \textit{HST/F547M} image. The contours of the warm H$_{2}$ molecular gas obtained from NaCo are shown in blue. \textbf{Bottom right:} \textit{Chandra} X-ray image in the 0.2-10 keV band. The contours of the warm H$_{2}$ molecular gas obtained from NaCo are shown in blue. colour scales are logarithmic and in arbitrary units. The FoV for all panels is 10 arcsec $\times$ 10 arcsec. North is up and east is to the left.}
 \label{n1386}
\end{figure*}

\begin{figure*}
 \includegraphics[width=0.85\textwidth]{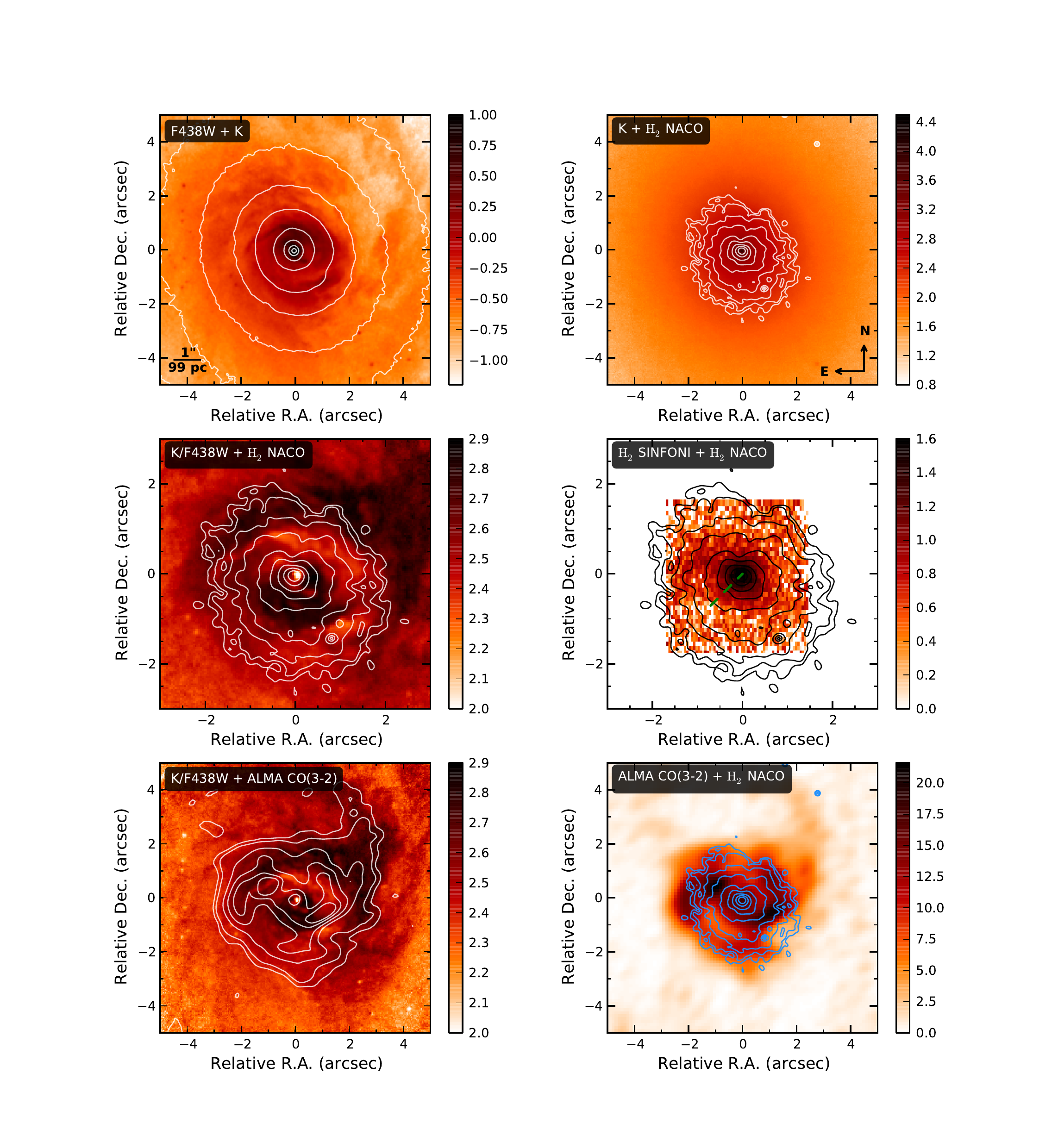}
 \protect\caption[n1566]{NGC\,1566. \textbf{Top left}: \textit{HST/F438W} image with NaCo/\textit{K}-band continuum contours in white. The FoV is 10 arcsec $\times$ 10 arcsec. \textbf{Top right:}NaCo/\textit{K}-band image with the same FoV as the previous panel. The contours of the warm H$_{2}$ molecular gas obtained from NaCo are shown in white. They start at three times the background noise level and increase in factors of 2. \textbf{Middle left:} \textit{K/F438W} ratio or dust map with a FoV of 6 arcsec $\times$ 6 arcsec. The contours of the warm H$_{2}$ molecular gas obtained from NaCo are shown in white. \textbf{Middle right:} 
Intensity map of the warm H$_{2}$ molecular gas obtained from IFU SINFONI. The FoV is the same as the previous panel. The contours of the warm H$_{2}$ molecular gas obtained from NaCo are shown in black. The orientation of the narrow line region (e.g. \citealt{1996ApJ...463..498S}) is indicated with a dashed green line. \textbf{Bottom left:} \textit{K/F438W} ratio or dust map with a FoV of 10 arcsec $\times$ 10 arcsec. The contours of the cold CO(3-2) molecular gas obtained from ALMA (\citealt{2014A&A...565A..97C}) are shown in white. Contours levels are 1.65, 2.7, 6.5, 10, 14 and 17 times 0.87 Jy beam$^{-1}$ km s$^{-1}$. \textbf{Bottom right:} ALMA CO(3-2) integrated intensity map (from \citealt{2014A&A...565A..97C}) with the same FoV as the previous panel. The contours of the warm H$_{2}$ molecular gas obtained from NaCo are shown in blue. The colour scale is linear in 0.87 Jy beam$^{-1}$ km s$^{-1}$. The colour palettes in the other panels are in arbitrary units. North is up and east is to the left.}
 \label{n1566}
\end{figure*}

\begin{figure*}
 \includegraphics[width=0.85\textwidth]{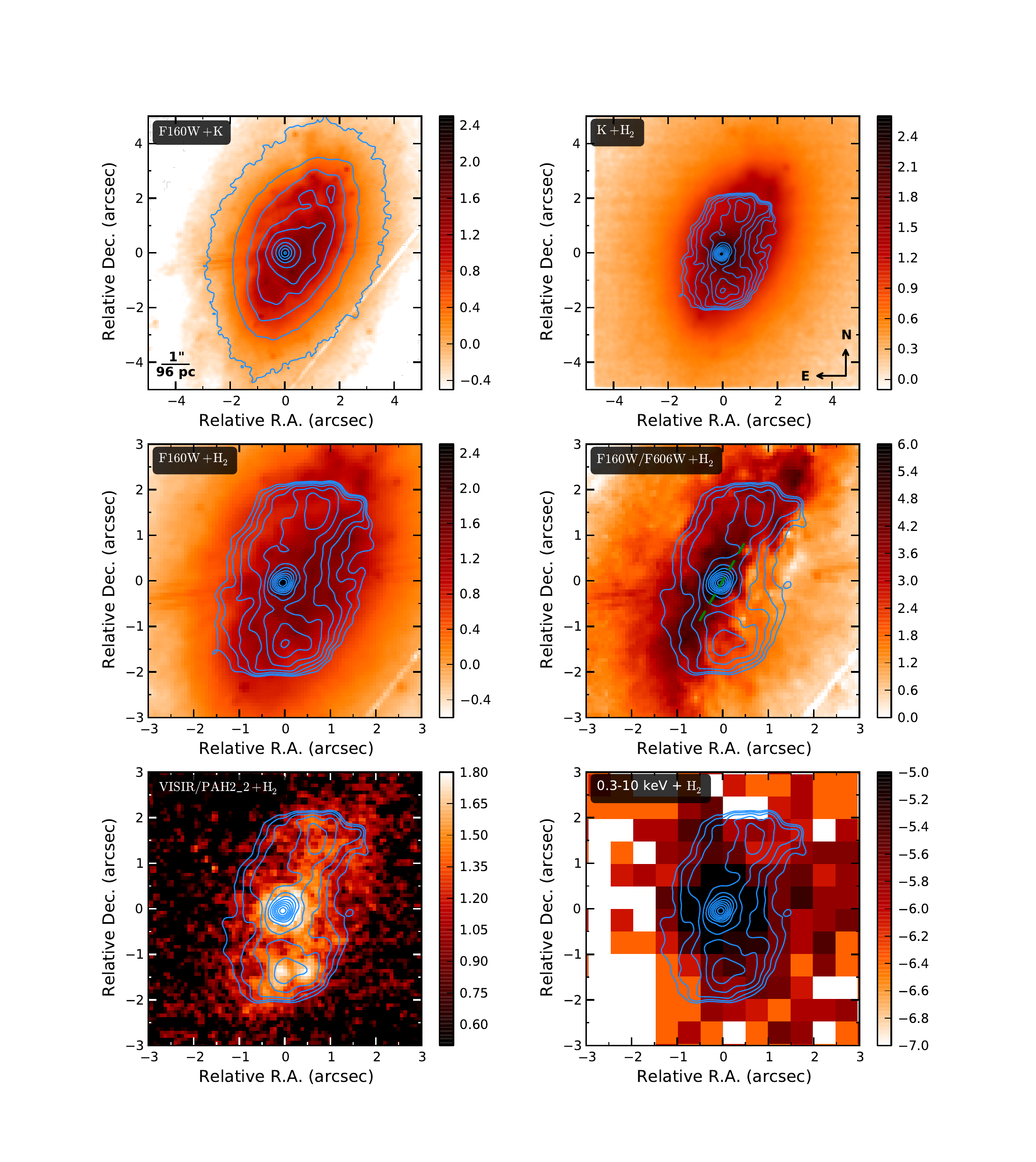}
 \protect\caption[n7582]{NGC\,7582. \textbf{Top left}: \textit{HST/F160W} image with NaCo/\textit{K}-band continuum contours in blue. The FoV is 5 arcsec $\times$ 5 arcsec. \textbf{Top right:}NaCo/\textit{K}-band image with the same FoV as the previous panel. The contours of the warm H$_{2}$ molecular gas obtained from IFU SINFONI are shown in blue. They start at three times the background noise level and increase in factors of 1.2. \textbf{Middle left:} \textit{HST/F160W} image with a FoV of 6 arcsec $\times$ 6 arcsec. The contours of the warm H$_{2}$ molecular gas obtained from IFU SINFONI are shown in blue. \textbf{Middle right:} \textit{F160W/F606W} ratio or dust map with the same FoV as the previous panel. The contours of the warm H$_{2}$ molecular gas obtained from IFU SINFONI are shown in blue. The orientation of the narrow line region (e.g. \citealt{1984ApJ...285..439U}; \citealt{1999A&AS..137..457M}) is indicated with a dashed green line. \textbf{Bottom left:} \textit{VISIR/PAH2\_2} image ($11.88\, \rm{\micron}$) with the same FoV as the previous panels. The contours of the warm H$_{2}$ molecular gas obtained from IFU SINFONI are shown in blue. \textbf{Bottom right:} \textit{Chandra} X-ray image in the 0.3-10 keV band with the same FoV as the previous panels. The contours of the warm H$_{2}$ molecular gas obtained from IFU SINFONI are shown in blue. colour scales are logarithmic and in arbitrary units. North is up and east is to the left.}
 \label{n7582}
\end{figure*}

\clearpage

\appendix
\cleardoublepage

\section{Discussion on individual objects} \label{individual}
\subsection*{CenA}
Centaurus A (also NGC5128 or CenA), at a distance of 3.8 $\pm$ 0.1 Mpc (\citealt{2010PASA...27..457H}), is the most nearby elliptical galaxy and a merger archetype. It hosts a Sy2 nucleus, covered by a prominent dust lane (see Fig.~\ref{cena}, top left), that emits a powerful radio and X-ray jet up to kpc scales. The NaCo narrow-band imaging of CenA shows that the molecular H$_{2}$ gas is concentrated in a spherical nuclear region of radius $\sim$0.3 arcsec ($\sim$5 pc; Fig.~\ref{cena}, top right panel). Further out, fainter shells of gas emission are traced up to $\sim$3 arcsec ($\sim$50 pc) north-east and south of the nucleus. This extended morphology is in agreement with the overall distribution observed in the SINFONI map (Fig.~\ref{cena}, middle-right panel), where shell features are observed (\citealt{2007ApJ...671.1329N}). These were interpreted as jet bow shocks, which is consistent with the orientation of the radio jet at pc scales (Fig.~\ref{cena}, middle-left panel). The warm H$_{2}$ gas distribution does not follow the filamentary dust morphology traced by the $K-I$ colour map nor the cold CO(2-1) gas emission (\citealt{2009ApJ...695..116E}; \citealt{2013ASPC..476...69E}; Fig.~\ref{cena}, middle and bottom left panels). The nucleus and jet of CenA are the most likely heating sources that excite the H$_{2}$ molecules (\citealt{2013ApJ...766...36B}).

\subsection*{Circinus}
\label{circinusdiscussion}
At a distance of 4.2 $\pm$ 0.8 Mpc (\citealt{1977A&A....55..445F}), Circinus is one of the
nearest Seyfert galaxies. It has a one-sided ionization cone north-west
of the nucleus (\citealt{1994A&A...291...18M}; see also Fig.~\ref{circinus}, middle-left panel). The nucleus of Circinus, which has
been resolved by near-IR imaging (\citealt{2004ApJ...614..135P}) and by
interferometric observations at mid-IR \citep{2014A&A...563A..82T}, is obscured
at optical wavelengths (\citealt{2004ApJ...614..135P}; Mezcua et al. in preparation).

The warm H$_2$ molecular gas emission is concentrated within a region of
$\sim$2 arcsec ($\sim$38 pc) radius, with the nuclear morphology in
north-east-south-west direction (perpendicular to the ionization
cone). This nuclear elongation is also observed in the small FoV of SINFONI presented by \cite{2006A&A...454..481M}; see also Fig.~\ref{circinus}, middle-right panel. 
The NaCo H$_2$ narrow-band image, with a larger FoV than SINFONI, reveals that extended H$_2$ emission in the form of $\sim$1 arcsec--wide arches is also detected $\sim$2 arcsec north-east and south-south-west of the nucleus. The northern spiral-like structure in the north-east curling turning towards west seems to be coincident with clumps of star formation in the $I$-band and H$\alpha$ image (Fig.~\ref{circinus}, bottom-left panel; Mezcua et al. in preparation), which suggests that the molecular gas in this region is heated by young stars, while the southern clumps seem to follow the spiral morphology of the dust colour map.

\subsection*{NGC~1068}
\label{ngc1068discussion}
The spiral SB galaxy NGC\,1068, at a distance of 14.1 Mpc (\citealt{1997A&A...320..399M}), hosts a very bright Sy2 nucleus that is obscured from optical wavelengths up to 1$\mu$m (e.g. \citealt{1997ApJ...476L..67C}; \citealt{2010MNRAS.402..724P,2014MNRAS.442.2145P}). The warm molecular H$_{2}$ gas distribution traced by NaCo covers a wider FoV than the SINFONI map and recovers the nuclear H$_{2}$ emission missing in the SINFONI map of \cite{2009ApJ...691..749M} (nuclear hole in Fig.~\ref{n1068}, upper right panel). It also shows clear diffraction effects (i.e. detector spikes) caused by the bright nucleus of this galaxy. Despite these diffraction effects and the nuclear differences, the extended NaCo morphology is consistent with that traced by SINFONI (e.g. \citealt{2009ApJ...691..749M}, \citealt{2014MNRAS.445.2353B}; see also \citealt{2002A&A...393...43G} for a VLT/ISAAC map): two prominent gas concentrations east and west of the nucleus and a collar-like structure south-west of it are observed on the NaCO map, similar to the SINFONI morphology and to the morphology of the slightly warm molecular gas ($T\sim$40--50 K) traced by recent CO(6-5) ALMA observations (\citealt{2014A&A...567A.125G}) and which extends up to 2 arcsec (140 pc) from the nucleus (see Fig.~\ref{n1068}, bottom-right panel).

 \cite{2009ApJ...691..749M} report the presence of a linear structure or northern tongue of H$_{2}$ emission streaming towards the nucleus. A good correlation between this northern tongue of H$_{2}$ gas and dust emission at 12$\mu$m was presented by these authors (also seen in the dust maps of \citealt{2014MNRAS.442.2145P}), which provides further evidence that the tongue is transporting feeding material to the AGN (\citealt{2006ApJ...646..774T}; though see \citealt{2014MNRAS.445.2353B} for an outflow scenario). The superposition of the warm H$_{2}$ gas contours on the $V-K$ colour dust map (Fig.~\ref{n1068}, middle-left panel) shows that the warm molecular gas is not only coincident with dust emission in the northern tongue but that an overall clear correlation is observed between the dust morphology and the H$_{2}$ emission, including the gas concentrations east and west of the nucleus and the ring-like structure $\sim$2 arcsec south of it. The correlation is however not one to one, as the warm H$_{2}$ gas presents a wider and more extended distribution than the dust. 
Despite the wider FoV of NaCo, no molecular H$_{2}$ gas emission is observed at scales larger than 2 arcsec, which indicates that all the warm molecular H$_{2}$ gas of NGC\,1068 is concentrated in the central 140 pc.

In Section~\ref{sourceh2} we found that the X-ray emission is the most likely cause of excitation of the nuclear warm H$_{2}$ molecular gas for most sources. In some regions close to the strong nuclear X-ray source and devoid of dust the X-ray emission could be able not only to excite the H$_{2}$ molecules but even to dissociate them. This extreme case might be observed for NGC\,1068, showcasing the role of dust in shielding H$_{2}$ molecules from the intense X-ray radiation: the south-west cavity seen in the dust map and H$_{2}$ contours (Fig.~\ref{n1068}, middle-left panel) coincides with an excess of \textit{V}-band (\textit{HST/F550W}; Fig.~\ref{n1068}, top-left panel) and \textit{Chandra} emission (Fig.~\ref{n1068}, middle-right panel), while the \textit{Chandra} knot $\sim$1 arcsec north of the nucleus coincides with a depression in the dust map and a hole in the H$_{2}$ molecular gas distribution (Fig.~\ref{n1068}, top-right panel); finally, an arch of X-ray \textit{Chandra} emission $\sim$0.5 arcsec east of the nucleus is observed to coincide with an equivalent depression zone in the dust and H$_{2}$ gas maps. For NGC\,1068 we might thus be seeing evidence that in the regions devoid of dust the H$_{2}$ molecule is not protected from the X-ray radiation but destroyed by it. This would explain the lack of H$_{2}$ emission in those regions where the X-ray emission is stronger, further supporting the existence of a clear link between H$_{2}$ molecular gas and X-ray radiation.

\subsection*{NGC~1097}
NGC\,1097 is a LINER/Sy1 galaxy located at $14.2\, \rm{Mpc}$ (\citealt{2009AJ....138..323T}) with a prototypical nuclear ring (diameter $\sim$18 arcsec; $\sim 1.2\, \rm{kpc}$). AO IR broad-band imaging by \cite{2005MNRAS.364L..28P} revealed dusty spirals around the nucleus and reaching the star-forming ring.

The H$_2$ image of NGC\,1097 (Fig.~\ref{n1097}, top-right panel) is dominated by the nuclear emission, which can be traced up to 2 arcsec ($\sim 140\, \rm{pc}$) from the nucleus. Further out, in the star-forming ring at $r \simeq$ 8 arcsec, several compact emission regions are detected. These H$_2$ point sources are coincident with the stellar clusters in the $K$-band (Fig.~\ref{n1097}, top-right panel; see also \citealt{2005MNRAS.364L..28P}). 

The nuclear H$_2$ emission shows a slightly elongated morphology along the east-west axis (middle-left panel in Fig.~\ref{n1097}). This is in contrast with the H$_2$ 1--0 S(1) intensity map obtained by \citet{2009ApJ...702..114D} with SINFONI (Fig.~\ref{n1097}, middle-right panel), which shows a more compact morphology elongated along the north-south direction. The discrepancy is probably ascribed to the method used to estimate the continuum emission in \citet{2009ApJ...702..114D}: the stellar and non-stellar continua are separated prior to the H$_2$ extraction, based on the dilution of the $2.3\, \rm{\micron}$ CO band head. In order to accommodate the differences between both maps, the non-stellar emission should be also more extended along the north-south axis, but this is not observed neither in the \textit{Ks}-band continuum nor in the $I - K$ colour map. The morphology of dust as traced by the $I - K$ colour (Fig.~\ref{n1097}, middle-left panel) does in any case not correlate with the nuclear extent of the warm H$_2$ emission.

\subsection*{NGC~1386}
NGC~1386 is a spiral Sb(s)a galaxy located in the Fornax cluster, at a distance of $15.3\, \rm{Mpc}$ (\citealt{2003ApJ...583..712J}), that hosts a Sy2 nucleus. NGC~1386 has a star-forming ring with a diameter of 30 arcsec ($2.2\, \rm{kpc}$) and tightly collimated ionized gas in the \textit{HST} {\oiii} and H$\alpha$ images extending 2 arcsec--3 arcsec towards north and south from the optical nucleus (\citealt{2000ApJS..128..139F}; \citealt{2014MNRAS.442.2145P}). It shows a one-sided jet oriented in the south-east direction (\citealt{1999ApJ...516...97N}; Fig.~\ref{n1386}, middle-left panel). The nucleus (or IR peak; \citealt{2014MNRAS.442.2145P}) is obscured, located 0.23 arcsec north of the optical peak of emission. 

The H$_2$ emission traced by NaCo presents a strong nuclear component and two extended regions up to 4 arcsec ($\sim$ 320 pc) towards the north and south of the nucleus (Fig.~\ref{n1386}, top-right panel). The latter were reported by \citet{2002MNRAS.331..154R} using long-slit spectroscopy in the near-IR. Its morphology, unveiled in this work for the first time, resembles that of the emission-line plumes detected in {\oiii} and {\nii}+H$\alpha$ by \cite{2000ApJS..128..139F}. In the $I - K$ colour map (Fig.~\ref{n1386}, middle-left panel), a dust lane is visible extending $\sim$ 3 arcsec towards north-east and south-west of the infrared nucleus. The peak of the H$_2$ emission is associated with the nucleus. The morphology of the extended H$_2$ emission resembles that of the dust lanes covering the IR nucleus, however it presents a much wider extent (middle-left panel in Fig.~\ref{n1386}). No H$_2$ emission is detected in the star-forming ring.

\subsection*{NGC~1566}
NGC~1566, at a distance of 20.5 Mpc (taken from NED adopting $H_0$ = 73 {\kms} Mpc$^{-1}$), is a nearly face-on SABbc spiral galaxy. 
It hosts a Seyfert nucleus that varies from type 1 to 2 (\citealt{1985ApJ...288..205A}) and has a one-sided ionization cone detected in {\oiii} that extends
0.3 arcsec towards south-east \citep{1996ApJ...463..498S}; see also Fig.~\ref{n1566}, middle-right panel. The optical \textit{HST} images
reveal dusty spirals reaching towards the nucleus (\citealt{1998ApJS..117...25M}; see also dust map in Fig.~\ref{n1566}, middle-left panel). Recent high-resolution CO ALMA observations show that the cold molecular CO(3-2) gas emission extends up to 3.5 arcsec forming a 2-arm nuclear spiral structure that corresponds very well with the dust lanes observed in the \textit{HST} images (\citealt{2014A&A...565A..97C}) and the dust map in Fig.~\ref{n1566} (bottom-left panel) . 

The NaCo warm H$_2$ molecular gas emission of NGC\,1566 reveals an almost circular diffuse morphology (Fig.~\ref{n1566}, top-right panel), slightly elongated in the east-west direction at $r<1$ arcsec, similar to the H$_2$ emission detected by SINFONI (Fig.~\ref{n1566}, middle-right panel). The extent of the warm H$_2$ gas region in the NaCo image is 2 arcsec ($\sim$220 pc), which is $\sim$2 times the size of that detected in the instrument-limited FoV of SINFONI and nearly the same as the extension of the H$_{2}$ emission detected by near-IR long-slit spectroscopy (\citealt{2002MNRAS.331..154R}). 
The warm H$_2$ gas emission does not follow the dust spiral structure as does the CO emission (Fig.~\ref{n1566}, bottom panels).

\subsection*{NGC~7582}\label{ngc7582discussion}
NGC\,7582 is a SBab galaxy located at $19.9\, \rm{Mpc}$ in the Grus Quartet \citep{2002A&A...393...57T}. Its Sy2 nucleus, a strong IR source only seen from the \textit{I}-band on (\citealt{2014MNRAS.442.2145P}), is surrounded by star-forming clusters resolved in the nuclear starburst (\citealt{2002ApJ...571L...7P}; also visible in the \textit{VISIR/PAH2\_2} image ($11.88\, \rm{\micron}$) in Fig.~\ref{n7582}). The morphology of the warm H$_2$ gas emission, traced by SINFONI, extends up to 2 arcsec (192 pc) and is associated with the starburst emission. The most embedded clusters, detected in the mid-IR and radio wavelengths \citep{2006MNRAS.369L..47W,2010MNRAS.401.2599O}, show a counterpart in the H$_2$ map (Fig.~\ref{n7582} bottom-left panel). A compact unresolved source of warm H$_2$ gas is detected at the position of the AGN and is most likely excited by X-ray emission (Fig.~\ref{n7582}, bottom-right panel).


\label{lastpage}

\end{document}